\documentclass[11pt]{article}
\usepackage[english]{babel}
\usepackage{supp/stylePreprint}

\title{Empirical comparison of win ratio and joint frailty models for recurrent event endpoints with applications in oncology and cardiology}

\shorttitle{Empirical comparison of win ratio and joint frailty models for recurrent event endpoints}

\date{}

\preprintauthors{%
\begin{adjustwidth}{-0.12\textwidth}{-0.12\textwidth}
  \setlength{\tabcolsep}{0pt}
  \begin{tabularx}{\linewidth}{@{}*{5}{>{\centering\arraybackslash}X}@{}}%
        \PreprintAuthorBox[0009-0003-4041-5016]{Adrien Oru\'e}[1] &
        \PreprintAuthorBox[0000-0003-1815-3016]{Derek Dinart}[2,3] &
        \PreprintAuthorBox[0000-0002-4975-9793]{Laurent Billot}[4] &
        \PreprintAuthorBox[0000-0001-7926-4671]{Carine Bellera}[2,3] &
        \PreprintAuthorBox[0000-0001-7109-4831]{Virginie Rondeau}[1,*]%
    \end{tabularx}%
\end{adjustwidth}%
}

\correspondingauthor{virginie.rondeau@u-bordeaux.fr}

\preprintaffiliations{%
\begin{adjustwidth}{-0.03\textwidth}{-0.03\textwidth}
    \setlength{\tabcolsep}{8pt}%
    \begin{tabularx}{\linewidth}{@{}*{4}{>{\arraybackslash}X}@{}}%
        \PreprintAffilBlock{1}{Department of Biostatistics, Bordeaux Population Health Research Center, INSERM U1219, Universit\'e de Bordeaux, Bordeaux, France} &
        \PreprintAffilBlock{2}{Clinical and Epidemiological Research Unit, Institut Bergoni\'e, Comprehensive Cancer Center, Inserm CIC1401, F-33000 Bordeaux, France} &
        \PreprintAffilBlock{3}{Univ. Bordeaux, Inserm, Bordeaux Population Health Research Center, Epicene team, UMR 1219, F-33000 Bordeaux, France} &
        \PreprintAffilBlock{4}{The George Institute for Global Health, Faculty of Medicine and Health, University of New South Wales, Sydney, Australia}%
    \end{tabularx}%
\end{adjustwidth}%
}

\hypersetup{
    pdftitle={Empirical comparison of win ratio and joint frailty models for recurrent event endpoints with applications in oncology and cardiology},
    pdfsubject={stat.AP, stat.CO, stat.ME},
    pdfauthor={Adrien Oru\'e, Derek Dinart, Laurent Billot, Carine Bellera, Virginie Rondeau},
    pdfkeywords={joint frailty models, win ratio, recurrent events, sample size, power, oncology, cardiology, randomized clinical trials},
}

\begin{document}

\maketitle

\begin{abstract}
  Composite endpoints that combine recurrent non-fatal events with a terminal event are increasingly used in randomized clinical trials, yet conventional time-to-first event analyses may obscure clinically relevant information. We compared two statistical frameworks tailored to such endpoints: the joint frailty model (JFM) and the last-event assisted recurrent-event win ratio (LWR). The JFM specifies proportional hazards for the recurrent and terminal events linked through a shared frailty, yielding covariate-adjusted, component-specific hazard ratios that account for informative recurrences and dependence with death. The LWR is a nonparametric, prioritized pairwise comparison that incorporates all observed events over follow-up and summarizes a population-level benefit of treatment while respecting a pre-specified hierarchy between death and recurrences. We first assessed the performance of the methods using simulations that varied both the gamma-frailty variance and the event rates. We next illustrated both approaches using two clinical application examples in oncology and cardiology, highlighting how conclusions depend on whether treatment primarily affects recurrent events, mortality, or both. The JFM provided component-specific estimates, while the LWR led to a summary measure of treatment effect with direction. Power was systematically improved with JFM, which thus appeared as the most reliable approach for inference and sample size estimation. Methodological extensions of the LWR to appropriately handle censoring and to formalize causal estimands remain a promising direction for future research.
\end{abstract}


\newpage

\section{Introduction}\label{sec1}

Clinical trials frequently use composite endpoints (CEs) to summarize treatment benefit when a disease manifests through multiple clinically relevant outcomes, to capture the burden of disease more comprehensively than any single component alone \cite{Freemantle2003-un,Armstrong2017-ma,Cordoba2010-tr}. A CE is typically defined as the occurrence of any event from a set of component outcomes (e.g., death, hospitalization, relapse). By incorporating multiple outcomes into a single criterion, one should expect an increase in the probability of observing events during the follow-up, potentially leading to a better statistical efficiency compared to a single component. In survival analysis, it is common to model CEs using a time-to-first-event approach: each participant contributes to a single event time corresponding to the first occurrence of any component of the CE, and treatment is assessed using Kaplan-Meier curves and Cox models \cite{Freemantle2003-un,Armstrong2017-ma}.

When using such time-to-first-event approaches, meaningful interpretation of the effect on the CE relies on strong hypotheses: clinical components should be of comparable clinical importance, occur with similar event rates, and the intervention may affect all components in the same direction \cite{McCoy2018-gd}. If these conditions are not satisfied, the composite outcome becomes a data-determined mixture weighted by component rates and relative effects, such that frequent but less serious events can dominate and obscure effects on more critical outcomes like mortality \cite{Montori2005-br,Ferreira-Gonzalez2007-fp,Cordoba2010-tr,Ferreira-Gonzalez2007-bo}. For instance, a CE consisting of death and relapse may favor control group if control group patients relapse early, while experimental group patients die later. Moreover, if a subject experienced a relapse then dies, only the time-to-relapse would be taken into account, hence not making use of the full history of events. In this context of time-to-first-event analysis, death is considered as a non-informative censoring.

We can also consider CEs consisting of a terminal event (e.g., death) and a lower-priority non-fatal component that can recur over follow-up (e.g., repeated hospitalizations or readmissions). In the present paper, we are interested in this type of CE. As before, a time-to-first-event analysis discards information after the first non-fatal event. In addition, we face three more challenges that together define the methodological problem addressed in this study. First, patients with multiple events induce within-subject dependence due to unobserved heterogeneity and state dependence (a history of events can increase the future risk of both non-fatal and fatal events). Second, once a terminal event occurs, it precludes any further recurrences, creating a semi-competing risks structure \cite{Liu2004-br,Rondeau2007-ef}. Finally, and unlike previously, the terminal event is usually informative: its timing is often strongly correlated with the recurrent event process. For example, patients with frequent early cancer relapses may be at higher risk of death \cite{Liu2004-br,Rondeau2007-ef}. Ignoring either the within-patient correlation or the dependency between processes can bias inference and reduce statistical power.

In this context of recurrent events and terminal event, we want to compare two methods that seem well suited to address these problems, but target two different estimands. The first one is the joint frailty model (JFM), which provides component-specific hazard ratios for the recurrent and terminal event processes, for each covariate of interest. Originally proposed in 2004 \cite{Liu2004-br} and extended in 2007 \cite{Rondeau2007-ef}, JFMs account for both unobserved heterogeneity and informative censoring by death through the joint modeling of recurrent events and the terminal event, linked via shared frailty. Model-based, it specifies hazard functions for the recurrent events and for the terminal event while introducing a shared patient-level frailty term to induce dependence between them \cite{Liu2004-br,Rondeau2007-ef}. This approach yields familiar estimands such as hazard ratios for treatment or covariate effects, which can be estimated semiparametrically (e.g., Breslow-type estimators \cite{Liu2004-br}, M-splines \cite{Krol2017-gs, Rondeau2007-ef}) or parametrically (e.g., Weibull \cite{Krol2017-gs}), and is implemented in the R package \texttt{frailtypack} \cite{Krol2017-gs}JFMs have been successfully applied in medical contexts like oncology to jointly model, for instance, tumor relapses and death \cite{Rondeau2007-ef}. A limitation is that the interpretation of random effects and subject-specific hazard ratios can be more nuanced than that of standard approaches such as a Cox model.

The second approach is the win ratio (WR), a nonparametric method based on pairwise comparison of outcomes, which has gained popularity as an alternative analysis for hierarchical CEs \cite{Pocock2012-jk,Pocock2024-fn}. Unlike the JFM, the WR yields a single population-level measure of treatment effect that summarizes the prioritized composite outcome as a whole, rather than estimating separate treatment effects for each individual component. Proposed by Pocock et al. in 2012 \cite{Pocock2012-jk}, it compares pairs of patients from the experimental and control arms across a hierarchy of outcomes, starting with the most clinically important (e.g., death) and proceeding to less critical outcomes (e.g., hospitalizations) if the first comparison is inconclusive. We have a "win" for the experimental arm if the treated patient has a better outcome than the control patient, a "loss" if the control patient fares better, and a "tie" if all comparisons are inconclusive \cite{Pocock2012-jk,Pocock2024-fn}. The standard WR statistic is then the number of wins divided by the number of losses, with an effect measure greater than 1 if the treatment is beneficial \cite{Pocock2012-jk,Pocock2024-fn}. The WR approach respects the hierarchical clinical importance of outcomes and makes use of all relevant patient data by allowing the comparison multiple outcomes, rather than only the first event \cite{Redfors2020-qd}. The FDA recommends this framework when hierarchical CEs are considered \cite{fda}. In 2022, Mao et al. \cite{Mao2022-qd} extended the WR to recurrent events by incorporating the total number and timing of recurrent events into the win/loss/tie calculation, implemented in their \texttt{WR} R package \cite{WR_package-rt}. This may increase statistical efficiency, often yielding higher power than a time-to-first or standard WR analysis when recurrent events are frequent \cite{Mao2022-qd}. Additionally, it remains a relatively recent methodology with which some investigators may still be unfamiliar.

From a trial-design perspective, planning sample size or power for CEs is often planned using standard event-based approximations for the logrank test or Cox model, such as Schoenfeld's formula \cite{schoenfeld_sample-size_1983}. These approaches rely on proportional hazards and non-informative right censoring and require specification of a single treatment effect for the whole CE. This can be difficult to justify, because component-specific treatment effects, component-specific incidences, and the dependence between the component event times jointly determine the marginal treatment effect on the composite \cite{Wu2012-zz}. This may induce non-proportional hazards even when proportional hazards holds for each component \cite{Cortes-Martinez2025-lo}. To overcome these limitations, Cort{\'e}s-Mart{\'i}nez et al. \cite{Cortes-Martinez2025-lo} developed a comprehensive R package \texttt{CompAREdesign}, which computes sample sizes and anticipated composite effect sizes (e.g., the geometric average hazard ratio) by modeling the joint distribution of two time-to-event components via copulas and marginal Weibull distributions. Still, although their framework still does not account for recurrent events. Recently, Dinart et al. \cite{dinart_sample_2024} proposed design procedures using JFMs, for power and sample size calculation that explicitly accounts for effects on recurrent events, death, or both. This seems naturally suited for CEs of recurrent events and death. Mao et al. \cite{Mao2022-zj} also derived formulas for power and sample size calculations using win ratio. However, even if it allows consideration of both the terminal and the non-terminal event, this applies only to a CE without recurrent events. In the present work, we proposed a modification of the \texttt{WR} package to implement a simulation-based sample size calculation using their framework, but adapted for recurrent events.

The objective of this work was to empirically compare JFMs and WR for CEs of recurrent events and death, with respect to (i) inferential performance (bias, type I error, and power), and (ii) sample size estimation. Depending on the frameworks, death has distinct statistical roles: (i) as an informative terminal event (JFMs), (ii) as the highest-priority clinical outcome (WR) and (iii) as a naive independent censoring mechanism (as in standard survival models). The remainder of the manuscript is organized as follows. In Section \ref{sec2}, we introduced the WR and JFM frameworks and the associated inference and design procedures. We reported results of our simulation study in Section \ref{sec3}. In Section \ref{sec4}, we applied our methods to real-world datasets from oncology and cardiology. We concluded in Section \ref{sec5} with a discussion and practical guidelines. To facilitate reproducibility, all code developed for this study is available in a public repository, including our modified \texttt{WR} package that implements simulation-based sample size calculation for recurrent events (see Section \ref{sec6}).

\section{Statistical methods}\label{sec2}

Within the win-ratio framework, death is treated as the highest-priority component of the composite endpoint, and pairwise comparisons are restricted to a common follow-up horizon. Within the joint frailty model, death is modeled as an informative terminal event that ends the recurrent-event process. Finally, we assumed that all the components of the CE were identified -- i.e., we had both event indicators and event times for all components.

\subsection{General principles regarding the win ratio framework}

This subsection concisely introduces the win ratio framework for standard composite endpoints (i.e., without recurrent events), as originally proposed by Pocock et al. \cite{Pocock2012-jk}, and as a basis for the extension to recurrent events described in Section 2.2.

\subsubsection{Pairing strategy}

Three main pairing strategies exist to form treated-control pairs for comparison \cite{Pocock2012-jk,Pocock2024-fn}: the matched-pair approach, the unmatched approach, and the stratified approach.

In the matched-pair strategy, a pre-specified prognostic risk score is first computed from baseline covariates -- typically using a Cox model excluding the treatment effect. Within each study arm, we then rank patients according to this score, and we create 1:1 cross-arm pairs by matching individuals with similar predicted risk. Patients in the larger arm who cannot be matched are dropped from the analysis. The principal strengths are improved comparability (as pairs have similar baseline prognoses), interpretability at the patient level ("like-with-like" pairing), and often greater efficiency when strong prognostic factors exist.

The unmatched approach generates all possible treated-control pairs without relying on a risk model or matching algorithm. Its strengths are maximal use of all randomized patients and broad applicability when credible risk scores are unavailable or disputed. However, because pairs are formed across the entire risk spectrum, many comparisons involve patients with markedly different baseline prognoses, which can dilute contrasts compared to risk-matched methods. Moreover, the induced dependence (each patient appears in many pairs, implying non-independent pairs) complicates variance estimation and hence hypothesis testing.

Finally, the stratified pairing strategy can be seen as an intermediate method that forms pairs only within pre-specified strata -- such as study center or yet discrete calendar-time windows -- and then aggregates the stratum-specific results across strata. Practically, this means creating all treated-control pairs within each stratum, allowing comparisons between patients who are more similar on the stratifying factors. This enables a form of "adjustment" on categorical covariates. The main strength of this approach is its ability to control for measured confounding through stratification on key variables. However, it has several limitations: it requires careful, prospective definition of strata; may suffer from reduced precision if strata are too narrow (resulting in few matched pairs); and can lead to limited contribution from strata with few informative pairings. Additionally, a robust method is needed to appropriately pool results across strata. Such pairing methods, however, need an appropriate way to pool results across strata, as originally detailed in the work by Dong et al. \cite{Dong2018-zd}.

In practice, the pairing choice should be driven by the estimand and by available baseline information. Because similar individuals are compare with each other, the matched approach can be more efficient and clinically interpretable. However, as it requires a predefined risk model and that some patients may be dropped from the analysis, researchers often favor the unmatched approach \cite{Redfors2020-qd,ajufo_fallacies_2023}. If one want to get something closer to the matched approach, the stratified approach can be a good compromise, as it allows to control for important prognostic factors without the need of a risk model and without dropping patients. Therefore, in the present work, we only consider the unmatched and stratified approaches.

\subsubsection{Comparison rule}

After forming pairs, we restrict the comparisons to a pair-specific follow-up horizon equal to the common observation window for the two subjects, namely the minimum of their observed follow-up times (i.e., the first censoring time due to death, loss to follow-up, or administrative). The comparison procedure is as follows:
\begin{itemize}
  \item First, compare the highest-priority component (e.g., death). The patient who experiences the event first is considered to have the worst outcome; otherwise, the result is inconclusive.
  \item If the first comparison is inconclusive (e.g., both patients are alive, or if deaths occur after their common shared time-horizon), we proceed to the next clinical component in the hierarchy and repeat the same comparison process
  \item We continue down the hierarchy until we identify a winner or all components have been compared inconclusively.
\end{itemize}
We declare the pair a "win" if the experimental patient has a better outcome than the control patient, a "loss" if the control patient has a better outcome, or a "tie" if all comparisons are inconclusive. This comparison rule respects the clinical priorities encoded in the hierarchy and ensures that more important clinical outcomes dominate the assessment of treatment benefit.


It is worth noticing that an unbalanced allocation ratio does not affect the pairing strategy nor the comparison rule. However, it may impact the total number of pairs formed, hence the win ratio estimate's precision.

\subsubsection{Estimand}

As originally stated in Pocock et al. \cite{Pocock2012-jk,Pocock2024-fn}, the standard win ratio statistic is the ratio between the number of wins $N_w$ and the number of losses $N_l$. Denoting $n_{\mathcal E}$ and $n_{\bar{\mathcal E}}$ as the number of patients in the experimental and control arms, respectively, for a total number of $n = n_{\mathcal E} + n_{\bar{\mathcal E}}$ subjects, the win ratio statistic is
\begin{align}
  \mathrm{WR}=\frac{N_w}{N_l}\in[0,\infty[,
\end{align}
with $\mathrm{WR}>1$ favoring the experimental arm. For instance, $\mathrm{WR}=2$ means that there are twice as many wins as losses.

Note that an alternative WR statistic exists, which is computed by dividing the proportion of wins by the proportion of losses. This alternative win ratio is
\begin{align}
  \mathrm{WR}^*=\frac{N_w\big/ n_{\mathcal E}n_{\bar{\mathcal E}}}{N_l\big/ n_{\mathcal E}n_{\bar{\mathcal E}}},
\end{align}
with "$N_w\big/ n_{\mathcal E}n_{\bar{\mathcal E}}$" the proportion of wins, and "$N_l\big/ n_{\mathcal E}n_{\bar{\mathcal E}}$" the proportion of losses. Both definitions are equivalent, as $\mathrm{WR}=\mathrm{WR}^*$. However, using win proportions is more convenient for both mathematical derivations and the reporting of results (e.g., 50\% of the comparisons made favor the treatment).

\begin{figure}[!h]
  \centering
  \includegraphics[width=0.7\linewidth]{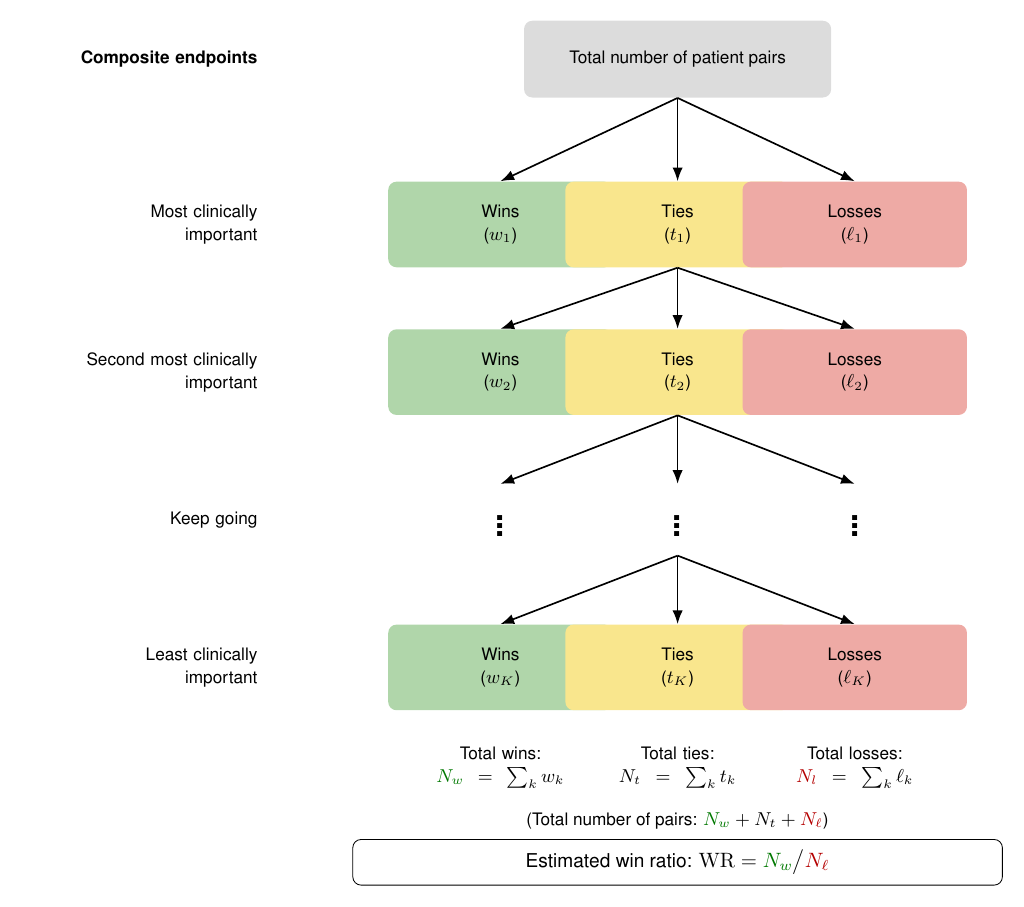}
  \caption{Illustration of the hierarchical pairwise comparison for a prioritized composite endpoint leading to the win ratio estimation.}
  \label{fig:wr_comp}
\end{figure}

An illustration of the hierarchical comparison that leads to the win ratio is given in Figure \ref{fig:wr_comp}, greatly inspired by the one of Treewaree et al. \cite{Treewaree2024-eo}.

\subsection{Win ratio for composite endpoint with a recurrent event}

When a lower-priority component can recur (e.g., repeated hospitalizations), one might consider using either the time to first recurrence or the total number of recurrences as a solution. However, the latter ignores timing and can induce ties if most counts are $0$ or $1$. Mao et al. proposed a method, using win functions, to incorporate both the frequency and timing of events \cite{Mao2022-qd}.

\subsubsection{Indexing and basic notation}
Let $\mathcal{E}=\{1,\dots,n_\mathcal{E}\}$ and $\bar{\mathcal{E}}=\{1,\dots,n_{\bar{\mathcal{E}}}\}$ denote the index sets for subjects in the experimental and control arms, respectively. For any subject $x\in \mathcal{E}\cup\bar{\mathcal{E}}$, let $D_x$ be the time to the terminal event (death), $N_{D,x}(t)=\mathbf{1}\{D_x\le t\}$ the death indicator, $N_{H,x}(t)$ the count of recurrent non-fatal events by time $t$, and $T_{j,x}$ the time of the $j$-th non-fatal event (when it exists). Let $C_x$ be the censoring time and $X_x=\min(C_x,D_x)$ the observed follow-up. For a treated-control pair $(e,\bar{e})\in\mathcal{E}\times\bar{\mathcal{E}}$, define the pair-specific shared follow-up horizon
$$
  \tau_{e, \bar{e}}=\min\bigl(X_e,X_{\bar{e}}\bigr).
$$
Finally, write $Y_x(t)=\{N_{D,x}(u), N_{H,x}(u): 0\le u\le t\}$ for the history observed up to time $t$.

\subsubsection{Win functions}

For a pair $(e,\bar{e})$ and horizon $\tau_{e, \bar{e}}$, the treated-win indicator under a given rule or win-function is $W(Y_e, Y_{\bar{e}};\tau_{e, \bar{e}})$. It is equal to 1 if the subject $e$ fares better than $\bar{e}$, and 0 otherwise. Likewise, we obtain the control-win indicator by swapping $e$ and $\bar{e}$. A "tie" correspond to both win-functions being zero. A total of four win-functions have been proposed by Mao et al. \cite{Mao2022-qd}.

The first win-function leads to the standard win ratio (SWR) that prioritizes survival and corresponds to the classic win ratio proposed by Pocock et al. \cite{Pocock2012-jk}. A treated-win occurs if the control dies earlier and before the shared horizon; otherwise, if both are alive at $\tau_{e, \bar{e}}$, the patient whose first recurrence occurs later wins. Formally:
$$
  \begin{aligned}
    W_{\mathrm{SWR}}(Y_e,Y_{\bar{e}}\ ;\tau_{e, \bar{e}})
     & =\mathbf{1}\bigl\{D_{\bar{e}}<D_e, D_{\bar{e}}\le \tau_{e, \bar{e}}\bigr\}+\mathbf{1}\bigl\{D_e>\tau_{e, \bar{e}}, D_{\bar{e}}>\tau_{e, \bar{e}}, T_{1,\bar{e}}<T_{1,e}, T_{1,\bar{e}}\le \tau_{e, \bar{e}}\bigr\}.
  \end{aligned}
$$
SWR uses only the first non-fatal event and thus may discard information from later recurrences.

The second win-function leads to the naive win ratio (NWR) that still prioritizes survival, but when both are alive at $\tau_{e, \bar{e}}$, it compares the counts of non-fatal events accrued by $\tau_{e, \bar{e}}$:
$$
  \begin{aligned}
    W_{\mathrm{NWR}}(Y_e,Y_{\bar{e}}\ ;\tau_{e, \bar{e}})
     & =\mathbf{1}\bigl\{D_{\bar{e}}<D_e, D_{\bar{e}}\le \tau_{e, \bar{e}}\bigr\}
    +\mathbf{1}\bigl\{D_e>\tau_{e, \bar{e}}, D_{\bar{e}}>\tau_{e, \bar{e}}, N_{H,\bar{e}}(\tau_{e, \bar{e}})<N_{H,e}(\tau_{e, \bar{e}})\bigr\}.
  \end{aligned}
$$
Once survival is deemed comparable, the winner is the individual with fewer recurrent events by $\tau_{e, \bar{e}}$. Because NWR ignores the timing of events, if both subjects have equal counts (especially when recurrences are rare), it results in many ties.

The third win-function leads to the first-event-assisted win ratio (FWR) that augments NWR with a rule to break ties based on time to first recurrence when the counts are equal and non-zero:
$$
  \begin{aligned}
    W_{\mathrm{FWR}}(Y_e,Y_{\bar{e}}\ ;\tau_{e, \bar{e}})
     & =W_{\mathrm{NWR}}(Y_e,Y_{\bar{e}}\ ;\tau_{e, \bar{e}})                                                                                                                                                           \\
     & \quad+\mathbf{1}\bigl\{D_e>\tau_{e, \bar{e}}, D_{\bar{e}}>\tau_{e, \bar{e}}, N_{H,\bar{e}}(\tau_{e, \bar{e}})=N_{H,e}(\tau_{e, \bar{e}})\ge 1, T_{1,\bar{e}}<T_{1,e}, T_{1,\bar{e}}\le \tau_{e, \bar{e}}\bigr\}.
  \end{aligned}
$$
If both patients survive to $\tau_{e, \bar{e}}$ with the same (non-zero) number of recurrences, it favors the one whose first event occurs later. FWR therefore emphasizes early differences in the onset of recurrence while maintaining NWR's count comparison as the primary secondary rule.

Finally, the last win-function leads to the last-event-assisted win ratio (LWR) augmenting NWR a rule to break ties based on time to last recurrence when the counts are equal and non-zero:
$$
  \begin{aligned}
    W_{\mathrm{LWR}}(Y_e,Y_{\bar{e}}\ ;\tau_{e, \bar{e}})
     & =W_{\mathrm{NWR}}(Y_e,Y_{\bar{e}}\ ;\tau_{e, \bar{e}})                                                                                                                                                            \\
     & \quad+\mathbf{1}\bigl\{D_e>\tau_{e, \bar{e}}, D_{\bar{e}}>\tau_{e, \bar{e}}, N_{H,e}(\tau_{e, \bar{e}})=N_{H,\bar{e}}(\tau_{e, \bar{e}})=k\ge 1,T_{k,\bar{e}}<T_{k,e}, T_{k,\bar{e}}\le \tau_{e, \bar{e}}\bigr\}.
  \end{aligned}
$$
When survival is comparable and event counts are tied, the winner is the one whose last event occurs later -- capturing the persistence of benefit across the trajectory of recurrences.

Table \ref{tab:win_functions} below summarizes the four win functions.
\begin{table}[h!]
  \centering
  \caption{Decision rules for the four win-function variants at a shared time horizon $\tau$.}
  \makebox[\textwidth][c]{%
    \small
    \begin{tabularx}{1.225\textwidth}{lXXX}
      \toprule
      Variant                              & Primary comparison & Secondary comparison          & Rule to break ties when counts are equal ($\ge 1$) \\
      \midrule
      Standard win ratio (SWR)             & Death time         & Time to first non-fatal event & ---                                                \\
      Naive win ratio (NWR)                & Death time         & \# of non-fatal events        & ---                                                \\
      First-event-assisted win ratio (FWR) & Death time         & \# of non-fatal events        & Time to first recurrence                           \\
      Last-event-assisted win ratio (LWR)  & Death time         & \# of non-fatal events        & Time of last recurrence                            \\
      \bottomrule
    \end{tabularx}
  }
  \label{tab:win_functions}
\end{table}

Since the LWR is recommended by Mao et al., we focused our analysis on this approach. In many cardiology and oncology settings, treatments may not only delay the first recurrence but also change the later course of the disease, by reducing the number of subsequent events or delaying them. The LWR accounts for all observed non-fatal events and uses the timing of the last recurrence to break ties when patients have the same number of events. This approach captures long-term differences in recurrence patterns that analyses based only on the first event or on event counts may miss. From now, "win ratio" refers to the recurrent-event win ratio based on the LWR, unless otherwise specified. In the same way, we will also only consider the LWR win function in the following, which will be noted as $W_{\mathrm{LWR}}(\cdot,\cdot) = W(\cdot, \cdot)$.

\subsubsection{Inference}

\paragraph{Notation and comparison horizon.} For a given win-function $W$ (e.g., LWR) satisfying the following regularity conditions \cite{Mao2022-qd}: for all pairs, (A1) we make comparisons at a common time-horizon and use only information observed up to that time; (A2) for any pair and horizon, win and loss indicators are mutually exclusive (i.e., we cannot have two wins or losses in one pair); and (A3) once death occurs for either subject, the win/loss status remains invariant thereafter. Particularly, (A3) implies that
$$
  W\!\bigl(Y_e,Y_{\bar{e}};\min(X_e,X_{\bar{e}})\bigr)=W\!\bigl(Y_e,Y_{\bar{e}};\min(C_e,C_{\bar{e}})\bigr),
$$
and likewise with $e$ and $\bar{e}$ swapped. This enables separation of outcomes from censoring in the asymptotic analysis.

With $n_{\mathcal{E}}$ treated patients and $n_{\bar{\mathcal{E}}}$ control patients, the win ratio estimator under rule $W$ is
\begin{align}
  \widehat{\mathrm{WR}}_{\mathrm{rec}}(W)=\frac{\frac{1}{n_{\mathcal{E}} n_{\bar{\mathcal{E}}}}\sum_{e\in\mathcal{E}}\sum_{\bar{e}\in\bar{\mathcal{E}}} W\!\bigl(Y_e,Y_{\bar{e}};\tau_{e,\bar{e}}\bigr)}{\frac{1}{n_{\mathcal{E}} n_{\bar{\mathcal{E}}}}\sum_{e\in\mathcal{E}}\sum_{\bar{e}\in\bar{\mathcal{E}}} W\!\bigl(Y_{\bar{e}},Y_e;\tau_{e,\bar{e}}\bigr)} =\frac{\hat w}{\hat \ell},
\end{align}
where $\hat w$ and $\hat \ell$ are the observed fractions of wins and losses, respectively.

\paragraph{Hypotheses.} Let $E$ be a random draw from $\mathcal{E}$ and $\bar{E}$ a random draw from $\bar{\mathcal{E}}$, independent of each other, and define
$w(t)=\mathbb{E}\!\left[W\bigl(Y_E,Y_{\bar{E}};t\bigr)\right]$ and $\ell(t)=\mathbb{E}\!\left[W\bigl(Y_{\bar{E}},Y_E;t\bigr)\right]$, the population win and loss probabilities when all patients are followed up to $\tau$, the end of follow-up. We test the null hypothesis of no treatment benefit, which we express as:
$$
  H_0:\;w(t)=\ell(t)\ \ \forall t\in[0,\tau],
$$
that corresponds to a win ratio equal to 1.

For the alternative, one may consider a beneficial treatment effect (i.e. a win ratio greater than 1), which we express as
$$
  H_A^{(1)}:\;w(t)\ge \ell(t)\ \ \forall t\in[0,\tau],\ \text{with strict inequality for some }t.
$$
Otherwise, one may simply consider the two-sided alternative (i.e. a win ratio different from 1), which we express as
\[
  \begin{aligned}
    H_A^{(2)}:\quad
                         & \Big\{ w(t)\ge \ell(t)\ \forall t\in[0,\tau],\
    \text{with strict inequality for some } t \Big\}                      \\
    \qquad \text{or}\ \  & \Big\{ w(t)\le \ell(t)\ \forall t\in[0,\tau],\
    \text{with strict inequality for some } t \Big\}.
  \end{aligned}
\]

Taken together, this testing framework is most naturally read as a test of stochastic dominance with respect to $W$ rather than a statement about a single point estimate of the win ratio. For instance, under the one-sided alternative $H_A^{(1)}$, a statistically significant result supports the conclusion that, over the entire follow-up window $[0,\tau]$, the subjects in the experimental arm tends to win more often than it loses. This implies that the treatment stochastically delays both recurrent events and the terminal event.

\paragraph{Test statistic.} Let us recall that $n = n_{\mathcal{E}}+n_{\bar{\mathcal{E}}}$ is the total number of subjects. The test statistic is hence
\begin{align}
  Z=\frac{\log\widehat{\mathrm{WR}}_{rec}(W)}{\hat\sigma(W)/\sqrt{n}},
\end{align}
where $\hat\sigma(W)$ is a consistent estimator of the standard deviation of the log win ratio, and is asymptotically standard normal under the null. Let $\alpha_1$ be the type-I error rate. Hence, for a two-sided test, reject $H_0$ if $|Z|>z_{1-\alpha_1/2}$; for a one-sided test toward benefit, reject $H_0$ if $Z>z_{1-\alpha_1}$. Details on the test statistic and variance estimator $\hat\sigma(W)$ are provided in Appendix \ref{app:wr_unstrat}.

\paragraph{Stratified inference.} To adjust for baseline heterogeneity, partition the sample into strata $s=1,\dots,S$ with weights $\omega^{(s)}=n^{(s)}/n$ where $n^{(s)}$ is the stratum size. Let $n_{\mathcal{E}}^{(s)}$ and $n_{\bar{\mathcal{E}}}^{(s)}$ be the experimental/control counts in stratum $s$. Compute the within-stratum win-loss fractions $\hat w^{(s)}$ and $\hat \ell^{(s)}$ similarly as before, using $\tau_{e,\bar{e}}^{(s)}=\min(X_e^{(s)},X_{\bar{e}}^{(s)})$. Then, the stratified win ratio is
\begin{align}
  \widehat{\mathrm{WR}}_{\mathrm{str}}=\frac{\sum_{s=1}^S \omega^{(s)}\hat w^{(s)}}{\sum_{s=1}^S \omega^{(s)}\hat \ell^{(s)}}.
\end{align}

Moreover, under the null hypothesis that the stratified win ratio equals 1 (i.e., no treatment benefit within every stratum), the Wald-type test is then
\begin{align}
  Z=\frac{\log\bigl(\widehat{\mathrm{WR}}_{\mathrm{str}}\bigr)}{\hat\sigma_{\mathrm{str}}(W)/\sqrt{n}},
\end{align}
which is asymptotically standard normal under the null. Based on our previous remark, it can also be seen as a test of stochastic dominance that holds within each stratum. The previous asymptotic results also naturally extend to the stratified framework; details are presented in Appendix \ref{app:wr_strat}. Mao et al. confirm tangible power gains from stratification \cite{Mao2022-qd}.

\paragraph{Several remarks.} First, win ratio depends on the censoring distribution of $\min(C_e,C_{\bar{e}})$ and is therefore not a censoring-robust effect measure in the sense of ICH E9(R1) \cite{Mao2022-qd}. Its use is primarily recommended for hypothesis testing. Second, FWR and LWR improve upon NWR by resolving ties in equal event counts using event timing, but they coincide with SWR when the non-fatal events are non-recurrent. When recurrences are frequent, recurrent-event win ratios (and especially LWR) have greater statistical power than SWR. Conversely, when recurrences are rare, NWR tends to produce more ties.

\subsubsection{Power and sample size calculation}

Existing approaches for sample size determination are limited to composite endpoints with two components and do not accommodate recurrent events. We nevertheless introduce the method of Mao et al. \cite{Mao2022-zj}, as we later compute a sample size under the simplifying assumption that the composite endpoint is non-recurrent.

Mao et al. \cite{Mao2022-zj} derived two sample size expressions: a nonparametric formula and a model-based formula. Since only the model-based formula is implemented in the \texttt{WR} package \cite{WR_package-rt}, we focus on presenting this formulation.
It is called "model-based" because they link the log-HR effect scale to the WR effect scale using a parametric model. For a given type-I risk $\alpha_1$, a power $1-\alpha_2$, and allocation ratio between experimental and control groups $q$, the required total sample size is
\begin{align}
  \label{eq:ss_wrrec}
  n=\frac{\zeta_0^2\bigl(z_{1-\alpha_1/2}+z_{1-\alpha_2}\bigr)^2}{q(1-q)\,(\delta_0^\top\xi)^2}.
\end{align}
A total of 3 parameters are needed to compute the sample size: $\zeta_0$, $\delta_0$, and $\xi$. $\zeta_0$, the standard rank deviation (SRD), corresponds to the standard deviation of the generalized rank $R(Y)$ -- a score that is positive when a treated–control comparison favors treatment, negative when it favors control, and zero otherwise -- under the null/reference distribution. In other words, under the study design (accrual, follow-up, censoring), $\zeta_0$ is the SRD of $R(Y)$ for the control arm. The second one, $\delta_0$, can be seen as a conversion factor from the log-HR scale to the WR scale. The last one, $\xi$, is the log-HR vector, that is, the log-HR for the recurrent event and the log-HR for the terminal event.

Note that both $\delta_0$ and $\zeta_0$ must be computed numerically, as they depend on study design factors (accrual, follow-up, censoring) and therefore cannot be determined directly. This estimation relies on Monte-Carlo procedures using the Gumbel-Hougaard copula with exponential margins, and is implemented in Mao et al.'s package \texttt{WR} \cite{WR_package-rt}.

\subsection{Joint frailty model}

Joint frailty models are the standard approach for analyzing recurrent event data in the presence of an absorbing terminal event, typically death, that may be informative for the recurrent process \cite{Liu2004-br,Rondeau2007-ef}. Thus, it naturally takes into account the intra-subject correlation through an individual random effect (the "frailty"), taken into account for both inference and sample size estimation.

\subsubsection{Model}

Consider subjects $i=1,\dots,n$ with recurrent events indexed by $j=1,\dots,n_i$. Under a calendar-time scale, the gamma-JFM specifies the conditional hazards for the recurrent event process and for the terminal event (death) as \cite{Liu2004-br}
$$
  \begin{cases}
    r_i(t \mid \omega_i)       & = \omega_i\, r_0(t)\,\exp\!\big(\beta_R^\top Z_i\big),                \\
    \lambda_i(t \mid \omega_i) & = \omega_i^{\alpha}\, \lambda_0(t)\,\exp\!\big(\beta_D^\top Z_i\big),
  \end{cases}
$$
where $r_0(\cdot)$ and $\lambda_0(\cdot)$ are baseline hazards, $Z_i$ denotes covariates, and $\beta_R,\beta_D$ are regression vectors (not necessarely equal). The shared random effect $\omega_i$ (individual frailty) has unit mean and variance $\theta$, with $\omega_i \overset{\text{i.i.d.}}{\sim} \Gamma(\text{shape}= \theta^{-1}, \text{scale}=\theta)$. The parameter $\alpha$ links the two processes ($\alpha>0$ implies that subjects at higher recurrent-event risk are also at higher risk of death, conditionally on covariates).

\subsubsection{Inference}

Two analytic time scales are common. The calendar scale measures time since study entry without resetting after events, capturing cumulative risk over follow-up; the gap scale resets after each recurrence, characterizing conditional inter-event risk. In this work, we focus on calendar time for estimation and design.

For subject $i$, let $C_i$ denote the non-terminal censoring time (administrative or loss to follow-up), $D_i$ the death time and $X_{ij}$ be the $j$-th inter-non-fatal-event time. Also define the observed time to the $j$-th outcome as $T_{ij}=\min\!\Big(\sum_{l=1}^{j} X_{il},\, C_i,\, D_i\Big)$, and the end of follow-up as $T_i^\star=\min(C_i, D_i)$. Let $\delta_{ij}$ be the indicator that the $j$-th recurrent event is observed, $\delta_i^\star=\mathbf{1}(D_i\le C_i)$ the death indicator at $T_i^\star$, and $n_i=\sum_j \delta_{ij}$ the total number of observed recurrences. Denote the cumulative baseline hazards by $R_0(t)=\int_0^t r_0(u)\,du$ and $\Lambda_0(t)=\int_0^t \lambda_0(u)\,du$. Considering right censoring and no left truncation, and after integrating out $\omega_i$, the marginal log-likelihood under the calendar-time formulation is
\begin{align*}
  \log(\mathcal{L}(\Theta))=\sum_{i=1}^n \Bigg[
                                           & \sum_{j=1}^{n_i}\delta_{ij}\Bigl(\log r_0(T_{ij})+\beta_R^\top Z_i\Bigr)
                                           +\delta_i^\star\Bigl(\log \lambda_0(T_i^\star)+\beta_D^\top Z_i\Bigr) -\log\!\Bigl(\Gamma(\tfrac{1}{\theta}) \,\theta^{1/\theta}\Bigr) \\
                                           & +\log\!\int_0^\infty
                                           \omega^{\sum_k \delta_{ik}+\alpha\,\delta_i^\star+1/\theta-1}\times
                                           \exp\!\Bigl(-\omega\bigl[\tfrac{1}{\theta}+e^{\beta_R^\top Z_i}R_0(T_i^\star)\bigr]
                                           -\omega^{\alpha}\,e^{\beta_D^\top Z_i}\Lambda_0(T_i^\star)\Bigr)\,d\omega
                                           \Bigg],
\end{align*}
where $\Theta=(\beta_R^\top,\beta_D^\top,\theta,\alpha)^\top$.

We model baseline risks $r_0,\lambda_0$ either semi-parametrically -- through splines or piecewise constant functions -- or parametrically -- under a Weibull distribution, for instance. These approaches are implemented in the \texttt{frailtypack} package \cite{Krol2017-gs}. If the two baseline hazards are modeled using cubic splines, as in the work by Rondeau et al. \cite{Rondeau2007-ef}, we define the following spline basis:
$$
  r_0(t)=\sum_{v=1}^{M_R} a_v\,M_v(t),\qquad
  \lambda_0(t)=\sum_{v=1}^{M_D} b_v\,M_v(t),
$$
where $M_R$ and $M_D$ are the number of basis functions for the recurrent and death processes, respectively; $a_v\ge 0$, $b_v\ge 0$ with $\sum_v a_v=1$ and $\sum_v b_v=1$; and $M_v(t)$ are the M-spline basis functions. Then $R_0(t)$ and $\Lambda_0(t)$ are linear combinations of the integrated basis. Then, one maximizes the following penalized log-likelihood
$$
  \log(\mathcal{L}_P(\Theta)) = \log(\mathcal{L}(\Theta))
  \;-\; \kappa_1 \int_0^\infty r_0"(t)^2\,dt
  \;-\; \kappa_2 \int_0^\infty \lambda_0"(t)^2\,dt .
$$
where $\kappa_1,\kappa_2$ are positive smoothing parameters controlling the roughness of the baseline hazards. These smoothing parameters can be chosen using cross-validation on smaller models (e.g., shared frailty model), as mentioned in this tutorial for using \texttt{frailtypack} \cite{Krol2017-gs}.

\subsubsection{Power and sample size calculation}

Here, we only present the method for the assessment of an exposure (i.e., treatment). For a complete discussion, see Dinart et al. \cite{dinart_sample_2024}.

Let $\Omega^\top=(\beta_R,\beta_D,r_0(\cdot),\lambda_0(\cdot),\theta,\alpha)$ be the full parameter vector. Denote $\psi$ one component of $\Omega$ (e.g., $\beta_R$ or $\beta_D$), of value $\psi_0$ under the null and $\psi_A$ under the alternative. Let consider the univariate test
$$
  H_0:\ \psi=\psi_0=0 \quad\text{vs}\quad H_A:\ \psi=\psi_A\neq 0,
$$
Under the null, the univariate Wald statistic
\begin{align}
  W = \frac{(\hat\psi-\psi_0)^2}{\widehat{\mathrm{Var}}(\hat\psi)}  \xrightarrow{d} \chi^2_1
\end{align}
follows a central chi-squared distribution, where $\widehat{\mathrm{Var}}(\hat\psi)\approx \bigl[n\widehat{\mathcal I}_1(\Omega)\bigr]^{-1}_{\psi,\psi}$. For power and sample size calculation, we needed this statistic distribution under $H_A$. The latter follows a noncentral $\chi^2_1(\mu)$ with noncentrality parameter $\mu>0$. We estimate the parameter $\mu$ algorithmically, and we approximate the per-subject Fisher information $\mathcal I_1(\Omega)=\mathbb E\!\left[U_i(\Omega)\,U_i(\Omega)^\top\right]$ via Monte Carlo simulations at the planning values $\Omega_A$ (i.e., under $H_A$), where $U_i(\Omega) = \partial_\Omega \ell_i(\Omega)$ and $\ell_i$ is the individual log-likelihood. Strictly speaking, the baseline hazards $r_0(\cdot)$ and $\lambda_0(\cdot)$ are estimated through a finite-dimensional parameterization (e.g., Weibull scale and shape parameters, or spline coefficients). Consequently, $\Omega$ is treated as a finite-dimensional vector of scalars, and $\partial_\Omega \ell_i(\Omega)$ denotes the partial derivatives with respect to all these scalar parameters. Although the baseline hazard parameters ($r_0, \lambda_0$) and frailty parameters ($\theta, \alpha$) are not the direct target of the hypothesis test, they act as nuisance parameters that contribute to the Fisher information matrix and the overall data-generating process. With simulated datasets $m=1,\dots,M$ having $n_m$ subjects, we estimate (thanks to the law of large numbers) $\hat{\mathcal I}_1$ with
$$
  \widehat{\mathcal I}_1(\Omega_A)=\frac{1}{\sum_{m=1}^M n_m}\sum_{m=1}^M\sum_{i=1}^{n_m} U_{im}(\Omega_A)\,U_{im}(\Omega_A)^\top.
$$

The estimation of the non-centrality parameter $\mu$ involves the following three-step procedure \cite{dinart_sample_2024}. Let $c = q_{Q,1-\alpha_1}$ be the $(1-\alpha_1)$-quantile of a central $\chi^2_Q$ given a specified type-I error $\alpha_1$. Hence, for target power $1-\alpha_2$ at level $\alpha_1$,
\begin{enumerate}
  \item Determine the $\alpha_1$-quantile $c$.
  \item Identify a non-centrality parameter $\vartheta$ satisfying: $\Pr\left(\chi^2_Q(\vartheta) > c\right) \geq 1 - \alpha_2$, where $1 - \alpha_2$ is the desired statistical power.
  \item Estimate the noncentrality parameter as
        $$
          \mu
          = \underset{x\in[0,\vartheta]}{\arg\min}\;
          \Big|\,\Pr\!\big(\chi^2_Q(x) > c\big) - (1-\alpha_2)\,\Big|
        $$
\end{enumerate}
In the univariate case (e.g., only testing for the exposure effect), $Q = 1$. Once $\mu$ is estimated, the required total sample size is
\begin{align}
  n \ge \mu\ \frac{\bigl[\widehat{\mathcal I}_1^{-1}(\Omega_A)\bigr]_{\psi,\psi}}{\psi_A^2}\,.
\end{align}
Equivalently, for a given $n$, compute
$$
  \mu(n) = n\frac{\psi_A^2}{\bigl[\widehat{\mathcal I}_1^{-1}(\Omega_A)\bigr]_{\psi,\psi}}
$$
and take power as $\Pr(\chi^2_1(\mu(n))>c)$.

For a multivariate linear constraint, let $H_0\!:C\,\Omega=0$ versus $H_A\!:C\,\Omega\neq 0$ with parameter vector $\Omega^\top=(\beta_R,\beta_D,r_0(\cdot),\lambda_0(\cdot),\theta,\alpha)$ and rank $Q=\mathrm{rank}(C)\ge1$. Under the null, the Wald statistic
$$
  W = n\,(C\,\hat\Omega)^\top \Bigl(C\,\widehat{\mathcal I}_1^{-1}(\Omega_A)\,C^\top\Bigr)^{-1}\!(C\,\hat\Omega)
  \;\xrightarrow{d}\; \chi^2_Q .
$$

As an illustration, to jointly test the treatment effects on both recurrence and death, consider
$$
  H_0:\ \{\beta_R=0\}\ \text{and}\ \{\beta_D=0\}
  \quad\text{vs.}\quad
  H_A:\ \{\beta_R\neq0\}\ \text{or}\ \{\beta_D\neq0\},
$$
so that $Q=2$ and
$$
  C\,\Omega
  =\begin{pmatrix}
    1 & 0 & 0 & 0 & 0 & 0 \\
    0 & 1 & 0 & 0 & 0 & 0
  \end{pmatrix}
  \!\!
  \begin{pmatrix}
    \beta_R & \beta_D & r_0(\cdot) & \lambda_0(\cdot) & \theta & \alpha
  \end{pmatrix}^\top
  =
  \begin{pmatrix}\beta_R\\ \beta_D\end{pmatrix}.
$$
The Wald statistic can be written as
\begin{align}
  W = n
  \begin{pmatrix}
    \hat\beta_R & \hat\beta_D
  \end{pmatrix}
  \Bigl(C\,\widehat{\mathcal I}_1^{-1}(\Omega_A)\,C^\top\Bigr)^{-1}
  \begin{pmatrix}
    \hat\beta_R \\ \hat\beta_D
  \end{pmatrix}
  \;\xrightarrow{d}\; \chi^2_2 .
\end{align}

\subsection{Statistical software}

The CRAN package \texttt{WR} \cite{WR_package-rt} (version 1.0) implements the four win-function variants (i.e., SWR, NWR, FWR, LWR) for inference through the \texttt{WRrec()} function. Sample size is computed with the \texttt{WRSS()} function, where the quantities $\zeta_0$, $w_0$, and $\delta_0$ are computed with \texttt{base()}.

For estimation based on joint models of recurrent and terminal events, the \texttt{frailtypack} package provides the function \texttt{frailtyPenal()} \cite{Krol2017-gs} (version 3.8.0). The \texttt{frailtyDesign} family of functions support the necessary sample size computations.

\section{Simulations studies}\label{sec3}

\subsection{Methods and settings}

We designed the simulation studies with two primary objectives: (i) to compare the performance of the JFM and WR (precisely, the LWR) for recurrent endpoints with a terminal event and (ii) to compare their empirical power.

Across all scenarios, we simulated a total of $K = 500$ datasets, each containing $N = 400$ subjects, evenly allocated between exposed and unexposed groups in a 1:1 ratio. Data generation relied on the inverse transform sampling method \cite{Bender2005-id} applied to the gamma-frailty JFM. For each dataset, $i = 1, \dots, N$ subjects were simulated following the procedure described in Rondeau et al. \cite{Rondeau2007-ef}. We imposed an administrative right-censoring time of $C = 3.0$ for all subjects. To ensure reproducibility, we set the \texttt{R} seed to $k=1,\dots,500$ (one seed per dataset). For each subject, we generated the following components:
\begin{itemize}
  \item An individual frailty term $u_i \sim \Gamma(\text{shape} = 1/\theta, \text{scale} = \theta)$ of unit mean and of variance $\theta$.
  \item A binary exposure covariate $Z_{1,i} \sim \text{Bernoulli}(0.5)$.
  \item Another binary covariate $Z_{2,i} \sim \text{Bernoulli}(0.5)$.
  \item A time to terminal event $T_i^{(D)}$, simulated from the hazard function $\lambda_i(t) = \lambda_0(t) u_i^{\alpha} \exp(\beta^{D} Z_{1,i})$.
  \item Recurrent event gap times $T_{ik}^{(R)}$, that we generated sequentially from the hazard function $r_i(t) = r_0(t) u_i \exp(\beta^{R}_1 Z_{1,i} + \beta^{R}_2 Z_{2,i})$. We generated recurrent events until $\sum_{l \leq j} T_{il}^{(R)} < \min(T_i^{(D)}, C)$, where $j$ denotes the current number of recurrent events.
\end{itemize}

For the baseline hazard functions, we considered a constant risk for both the terminal and recurrent events, with $r_0=3/2$ and $\lambda_0=1/2$. We also considered six scenarios to evaluate the impact of varying levels of heterogeneity, as well as differences in the incidence of recurrent and terminal events. First, we investigated three scenarios characterized by different levels of heterogeneity: standard heterogeneity ($\theta = 0.5$), near-zero heterogeneity ($\theta = 0.01$), and high heterogeneity ($\theta = 1$). Two additional scenarios were defined under $\theta = 0.5$: (i) a high terminal event rate combined with a low recurrent event rate ($\lambda_0 = 2$, $r_0 = 1/5$), and (ii) a high recurrent event rate combined with a low terminal event rate ($r_0 = 2$, $\lambda_0 = 1/7$). Finally, to compare the two methods under misspecification (for the JFM), we defined a scenario with log-logistic baseline hazards: recurrent events with $\text{(shape, scale)} = (1.4, 2.0)$ and death with $\text{(shape, scale)} = (1.2, 3.0)$. Under this parameterization, the baseline median equals the scale, giving a median recurrence gap of 2.0 time-units and a median survival of 3.0. The shapes ($\ge 1$) yield unimodal hazards. Across all scenarios, regression coefficients were fixed at $\beta^{D} = \log(0.8)$ for the terminal event and $\beta^{R}_1 = \log(0.7), \beta^{R}_2=\log(0.9)$ for recurrent events. The association parameter was fixed at $\alpha = 1$ and was not estimated. The pseudocode for the data generation is provided in Appendix \ref{app:gen_proc}.

All analyses were performed in \texttt{R} (version 4.5.2) on a Dell Pro Max 16. If any issues arise during analyses, they will be reported following Wunsch et al.'s error handling workflow \cite{Wunsch2025-eg}. We fitted gamma-joint frailty models using the \texttt{frailtyPenal()} function from the \texttt{frailtypack} package \cite{Krol2017-gs} on version 3.8.0, under a parametric specification with a Weibull baseline hazard. Estimation was initiated with 50 quadrature nodes; when numerical issues occurred (e.g., non-convergence or likelihood divergence), models were re-estimated with 32 nodes and, if required, with 20 nodes. A maximum of 100 iterations was specified as the convergence criterion for each model-fitting procedure. A model is considered "fitted" if no numerical issues arose during estimation. On the other hand, as the inverse transform sampling for the JFM does not allow direct specification of the "true win ratio", we obtained this value empirically for each scenario. To overcome this limitation, we proceeded as follows. For each scenario, we generated, using the same data-generation procedure as above, 50 datasets, each containing $100,000$ individuals (i.e., $50,000$ simulated subjects in each of the exposure arms), and estimated the LWR (both unstratified and stratified on $Z_2$) using a computationally optimized version of Mao et al.'s \texttt{WR} R package \cite{WR_package-rt}. The resulting estimate was considered an "asymptotic" true value of the win ratio and was subsequently used for the assessment of bias and coverage rates in each simulation scenario.

We estimated empirical power as the proportion of simulated replicates in which the null hypothesis was rejected at the prespecified two-sided significance level $\alpha_1 = 5\%$. For the JFM, we tested $H_0: \beta^{R}_1=0 \text{ and } \beta^{D}=0 \text{ vs. } H_1: \beta^{R}_1\neq 0  \text{ or } \beta^{D}\neq 0$, with a multivariate Wald test (2-df). For the WR analysis, we tested $H_0:\ WR=1 \text{ vs. } H_1: WR\neq 1$.

\subsection{Results}

\subsubsection{Data summary}

Table \ref{tab:sim_summary} summarizes the main characteristics of the simulated datasets across the six scenarios. Across the 3,000 generated datasets (6 scenarios $\times$ 500 replicates), the overall death rate averaged 59.85\% (dataset range 22.5\%-96.0\%). Because administrative right-censoring at $C=3.0$ was the only censoring mechanism, the administrative censoring rate corresponds to the proportion of subjects who did not experience the terminal event by the end of follow-up, which equals $1-\Pr(T^{(D)}\le C)$ and therefore averaged 40.15\% (range 4.0\% - 77.5\%). Subjects experienced on average 1.6 recurrent events, with means ranging from 0.08 (scenario 4) to 3.73 (scenario 5) events and with individual counts ranging from 0 to 40 (40 being observed in scenario 5; otherwise $\leq 24$). On average, 45.2\% of subjects experienced no recurrent event (range: 10.0\%-96.0\%).

\begin{table}[ht!]
  \centering
  \caption{Summary characteristics of simulated datasets across the six scenarios.}
  \label{tab:sim_summary}
  \renewcommand{\arraystretch}{1.2}
  \makebox[\textwidth][c]{%
    \resizebox{1.2\textwidth}{!}{%
      \begin{tabular}{@{}l rrr rrr rrr@{}}
        \toprule
                                                                     & \multicolumn{3}{c}{\makebox[0pt][c]{\textbf{Death rate (\%)}}}
                                                                     & \multicolumn{3}{c}{\makebox[0pt][c]{\textbf{Rec. per subjects}}}
                                                                     & \multicolumn{3}{c}{\makebox[0pt][c]{\textbf{Zero event rate (\%)}}}                                                                                                                              \\
        \cmidrule(lr){2-4} \cmidrule(lr){5-7} \cmidrule(lr){8-10}
        \textbf{Scenario}                                            & Mean                                                                & SD   & Range          & Mean & SD   & Range               & Mean  & SD              & Range                                \\
        \midrule
        Scenario 1 -- Base case                                      & 64.27                                                               & 2.34 & 56.0\,--\,71.8 & 1.72 & 0.10 & 0\,--\,22           & 8.45  & \phantom{1}1.41 & 4.75\,--\,12.25                      \\
        Scenario 2 -- Low heterogeneity                              & 73.62                                                               & 2.31 & 66.0\,--\,79.3 & 1.97 & 0.10 & 0\,--\,14           & 1.03  & 10.50           & \phantom{1}0.00\,--\,\phantom{1}2.50 \\
        Scenario 3 -- High heterogeneity                             & 57.35                                                               & 2.48 & 51.3\,--\,65.3 & 1.54 & 0.10 & 0\,--\,24           & 16.94 & \phantom{1}1.83 & 11.25\,--\,21.75                     \\
        Scenario 4 -- High death rate, low nonfatal event rate       & 92.65                                                               & 1.32 & 89.3\,--\,96.0 & 0.08 & 0.02 & 0\,--\,\phantom{2}4 & 6.49  & \phantom{1}1.27 & \phantom{1}3.25\,--\,10.25           \\
        Scenario 5 -- High nonfatal event rate, low death rate       & 29.53                                                               & 2.20 & 22.5\,--\,36.0 & 3.73 & 0.19 & 0\,--\,40           & 7.97  & \phantom{1}1.32 & \phantom{1}4.25\,--\,12.25           \\
        Scenario 6 -- Misspecification (loglogistic baseline hazard) & 41.66                                                               & 2.53 & 32.0\,--\,48.5 & 0.53 & 0.04 & 0\,--\,\phantom{2}9 & 34.02 & \phantom{1}2.37 & 27.75\,--\,40.25                     \\
        \bottomrule
      \end{tabular}
    }
  }
  \begin{tablenotes}
    \item \textbf{Notes.} Summary statistics are computed across the 500 simulated datasets per scenario. "Zero event rate (\%)" denotes the percentage of simulated subjects that did not experience any events, both terminal and nonterminal (only one observation, which corresponds to a censoring, is counted).
    \item \textbf{Abbreviations.} Rec., recurrences; SD, standard deviation.
  \end{tablenotes}
\end{table}

\subsubsection{Estimations}

\begin{table}[ht!]
  \centering
  \caption{Simulation performance of the Joint Frailty Model across scenarios}
  \label{tab:jfm_all}
  \renewcommand{\arraystretch}{1.2}
  \makebox[\textwidth][c]{%
    \resizebox{1.225\textwidth}{!}{%
      \begin{tabular}{lrrrrrrr}
        \toprule
        \textbf{Parameter} & \textbf{True value} & \textbf{Mean estimate} & \textbf{Abs. bias} & \textbf{Rel. bias (\%)} & \textbf{Empirical SE} & \textbf{Asymptotic SE} & \textbf{Coverage (\%)} \\
        \midrule
        \multicolumn{8}{l}{\textbf{Scenario 1 -- base ($\alpha=1$, $\theta=0.5$, $r_0=3/2$ and $\lambda_0=1/2$)}}                                                                                  \\
        \multicolumn{8}{l}{$\quad$\ \emph{Power = 0.820 ; MCSE = 0.017 ; IC MC 95\% = (0.783, 0.853) ; Models fitted = 498/500}}                                                                   \\
        $\beta^{R}_1$      & -0.3567             & -0.3570                & -0.0003            & -0.1000                 & 0.1168                & 0.1112                 & 93.98                  \\
        $\beta^{R}_2$      & -0.1054             & -0.1104                & -0.0050            & -4.7900                 & 0.1068                & 0.1018                 & 93.98                  \\
        $\beta^{D}$        & -0.2231             & -0.2229                & 0.0002             & 0.1200                  & 0.1456                & 0.1488                 & 93.98                  \\
        $\theta$           & 0.5000              & 0.4736                 & -0.0264            & -5.2700                 & 0.1324                & 0.0698                 & 87.15                  \\
        \midrule
        \multicolumn{8}{l}{\textbf{Scenario 2 -- low heterogeneity ($\theta=0.01$)}}                                                                                                               \\
        \multicolumn{8}{l}{$\quad$\ \emph{Power = 0.978 ; MCSE = 0.007; IC MC 95\% = (0.961, 0.989); Models fitted = 490/500}}                                                                     \\
        $\beta^{R}_1$      & -0.3567             & -0.3608                & -0.0041            & -1.1500                 & 0.0715                & 0.0772                 & 95.51                  \\
        $\beta^{R}_2$      & -0.1054             & -0.1069                & -0.0016            & -1.5000                 & 0.0698                & 0.0765                 & 96.12                  \\
        $\beta^{D}$        & -0.2231             & -0.2254                & -0.0022            & -1.0000                 & 0.1239                & 0.1203                 & 93.88                  \\
        $\theta$           & 0.0100              & 0.0552                 & 0.0452             & 451.6800                & 0.0280                & 0.0109                 & 19.80                  \\
        \midrule
        \multicolumn{8}{l}{\textbf{Scenario 3 -- high heterogeneity ($\theta=1$)}}                                                                                                                 \\
        \multicolumn{8}{l}{$\quad$\ \emph{Power = 0.692 ; MCSE = 0.021 ; IC MC 95\% = (0.649, 0.732) ; Models fitted = 500/500}}                                                                   \\
        $\beta^{R}_1$      & -0.3567             & -0.3719                & -0.0152            & -4.2600                 & 0.1338                & 0.1345                 & 93.60                  \\
        $\beta^{R}_2$      & -0.1054             & -0.1047                & 0.0007             & 0.6300                  & 0.1201                & 0.1153                 & 94.40                  \\
        $\beta^{D}$        & -0.2231             & -0.2324                & -0.0093            & -4.1600                 & 0.1742                & 0.1706                 & 94.00                  \\
        $\theta$           & 1.0000              & 0.8626                 & -0.1374            & -13.7400                & 0.2726                & 0.0876                 & 86.00                  \\
        \midrule
        \multicolumn{8}{l}{\textbf{Scenario 4 -- high death rate, low nonfatal event rate ($r_0=1/5$, $\lambda_0=2$)}}                                                                             \\
        \multicolumn{8}{l}{$\quad$\ \emph{Power = 0.344 ; MCSE = 0.021 ; IC MC 95\% = (0.302, 0.387) ; Models fitted = 497/500}}                                                                   \\
        $\beta^{R}_1$      & -0.3567             & -0.2941                & 0.0625             & 17.5300                 & 0.3742                & 0.3627                 & 94.57                  \\
        $\beta^{R}_2$      & -0.1054             & -0.1089                & -0.0035            & -3.3700                 & 0.3718                & 0.3613                 & 96.18                  \\
        $\beta^{D}$        & -0.2231             & -0.1680                & 0.0551             & 24.7000                 & 0.1095                & 0.1069                 & 91.75                  \\
        $\theta$           & 0.5000              & 0.0417                 & -0.4583            & -91.6700                & 0.1236                & 0.0109                 & \phantom{9}5.63        \\
        \midrule
        \multicolumn{8}{l}{\textbf{Scenario 5 -- high nonfatal event rate, low death rate ($r_0=2$, $\lambda_0=1/7$)}}                                                                             \\
        \multicolumn{8}{l}{$\quad$\ \emph{Power = 0.932 ; MCSE = 0.11 ; IC MC 95\% = (0.906, 0.952) ; Models fitted = 498/500}}                                                                    \\
        $\beta^{R}_1$      & -0.3567             & -0.3490                & 0.0077             & 2.1500                  & 0.0922                & 0.0903                 & 94.78                  \\
        $\beta^{R}_2$      & -0.1054             & -0.1057                & -0.0003            & -0.3000                 & 0.0896                & 0.0859                 & 94.18                  \\
        $\beta^{D}$        & -0.2231             & -0.2251                & -0.0020            & -0.8800                 & 0.2100                & 0.1999                 & 94.38                  \\
        $\theta$           & 0.5000              & 0.4826                 & -0.0174            & -3.4700                 & 0.1001                & 0.0555                 & 91.77                  \\
        \midrule
        \multicolumn{8}{l}{\textbf{Scenario 6 -- Misspecification (loglogistic baseline hazard)}}                                                                                                  \\
        \multicolumn{8}{l}{$\quad$\ \emph{Power = 0.562 ; MCSE = 0.22 ; IC MC 95\% = (0.517, 0.606) ; Models fitted = 469/500}}                                                                    \\
        $\beta^{R}_1$      & -0.3567             & -0.3393                & 0.0173             & 4.8600                  & 0.1699                & 0.1505                 & 90.41                  \\
        $\beta^{R}_2$      & -0.1054             & -0.1096                & -0.0043            & -4.0700                 & 0.1435                & 0.1472                 & 95.74                  \\
        $\beta^{D}$        & -0.2231             & -0.2123                & 0.0109             & 4.8800                  & 0.1710                & 0.1652                 & 93.82                  \\
        $\theta$           & 0.5000              & 0.2422                 & -0.2578            & -51.5600                & 0.2901                & 0.0636                 & 38.17                  \\
        \bottomrule
      \end{tabular}
    }
  }
  \begin{tablenotes}
    \item \textbf{Abbreviations.} SE, standard error; MCSE, Monte Carlo standard error; IC MC 95\%, Monte Carlo 95\% confidence interval of the power estimate
  \end{tablenotes}
\end{table}

\begin{table}[!htbp]
  \centering
  \caption{Simulation performance of the last-event assisted win ratio across scenarios}
  \label{tab:lwr_all}
  \renewcommand{\arraystretch}{1.2}
  \makebox[\textwidth][c]{%
    \resizebox{1.25\textwidth}{!}{%
      \begin{tabular}{lrrrrrrrrr}
        \toprule
        \textbf{Type}      & \textbf{"True" value} & \textbf{Mean WR} & \textbf{Abs. bias} & \textbf{Rel. bias (\%)} & \textbf{Mean $1/\mathrm{WR}$} & \textbf{Emp. SE} & \textbf{Asymp. SE} & \textbf{Coverage (\%)} & \textbf{Power (\%)} \\
        \midrule
        \multicolumn{10}{l}{\textbf{Scenario 1 -- base ($\alpha = 1$, $\theta = 0.5$, $r_0 = 1.5$, $\lambda_0 = 0.5$)}}                                                                                                                     \\
        \quad Unstratified & 1.2253                & 1.2325           & 0.0072             & 0.5877                  & 0.8218                        & 0.1388           & 0.1445             & 95.6                   & 40.6                \\
        \quad Stratified   & 1.2254                & 1.2328           & 0.0074             & 0.6057                  & 0.8216                        & 0.1388           & 0.1447             & 95.8                   & 40.6                \\
        \midrule
        \multicolumn{10}{l}{\textbf{Scenario 2 -- low heterogeneity ($\theta = 0.01$)}}                                                                                                                                                     \\
        \quad Unstratified & 1.2901                & 1.3019           & 0.0118             & 0.9153                  & 0.7795                        & 0.1576           & 0.1521             & 93.8                   & 59.2                \\
        \quad Stratified   & 1.2904                & 1.3015           & 0.0111             & 0.8615                  & 0.7797                        & 0.1575           & 0.1523             & 94.0                   & 58.8                \\
        \midrule
        \multicolumn{10}{l}{\textbf{Scenario 3 -- high heterogeneity ($\theta = 1.0$)}}                                                                                                                                                     \\
        \quad Unstratified & 1.1912                & 1.2137           & 0.0225             & 1.8849                  & 0.8355                        & 0.1461           & 0.1450             & 94.2                   & 31.4                \\
        \quad Stratified   & 1.1912                & 1.2138           & 0.0226             & 1.8991                  & 0.8353                        & 0.1459           & 0.1451             & 93.8                   & 31.0                \\
        \midrule
        \multicolumn{10}{l}{\textbf{Scenario 4 -- Low nonfatal event rate, high death rate ($r_0=1/5$, $\lambda_0=2$)}}                                                                                                                     \\
        \quad Unstratified & 1.1962                & 1.2068           & 0.0106             & 0.8862                  & 0.8393                        & 0.1373           & 0.1410             & 96.4                   & 34.6                \\
        \quad Stratified   & 1.1962                & 1.2073           & 0.0111             & 0.9286                  & 0.8390                        & 0.1380           & 0.1413             & 96.0                   & 35.0                \\
        \midrule
        \multicolumn{10}{l}{\textbf{Scenario 5 -- High nonfatal event rate, low death rate ($r_0=2$, $\lambda_0=1/7$)}}                                                                                                                     \\
        \quad Unstratified & 1.3465                & 1.3555           & 0.0090             & 0.6672                  & 0.7489                        & 0.1669           & 0.1600             & 94.4                   & 70.8                \\
        \quad Stratified   & 1.3470                & 1.3564           & 0.0094             & 0.6979                  & 0.7484                        & 0.1673           & 0.1603             & 94.0                   & 70.8                \\
        \midrule
        \multicolumn{10}{l}{\textbf{Scenario 6 -- Misspecification (loglogistic baseline hazard)}}                                                                                                                                          \\
        \quad Unstratified & 1.2514                & 1.2558           & 0.0044             & 0.3551                  & 0.8107                        & 0.1693           & 0.1621             & 93.8                   & 40.4                \\
        \quad Stratified   & 1.2515                & 1.2563           & 0.0048             & 0.3797                  & 0.8105                        & 0.1699           & 0.1622             & 93.2                   & 40.6                \\
        \bottomrule
      \end{tabular}%
    } 
  } 
  \begin{tablenotes}
    \item \textbf{Abbreviations.} SE, standard error; WR, win ratio.; Emp., empirical; Asymp., asymptotic.
  \end{tablenotes}

\end{table}

Across scenarios, point estimation was well calibrated for both approaches (Tables \ref{tab:jfm_all}-\ref{tab:lwr_all}). For the JFM, regression coefficients for the recurrent process and the terminal event were essentially unbiased, with relative biases typically within $\pm$5\% (except for scenario 4) and empirical standard errors closely matching asymptotic standard errors. The same pattern held for the LWR estimates, where empirical and asymptotic standard errors were well aligned and 95\% interval coverage remained close to nominal ($\approx$93-96\% across scenarios). In contrast, the frailty variance $\theta$ was consistently the most challenging parameter: under near-homogeneity (Scenario 2) the mean estimate was 0.055 with substantial under-coverage (20\%), and under baseline-hazard misspecification (Scenario 6) $\theta$ was markedly underestimated (mean 0.242 vs. 0.5; coverage 38\%). The high death rate/low non-fatal event setting (Scenario 4) produced the most extreme under-coverage for $\theta$ (6\%). Model fitting for the JFM was near-perfect in most settings but deteriorated under misspecification (469/500 fits). When convergence issues arose, they were due to either likelihood blow-up or failure to converge within the maximum number of iterations. Overall, no runtime issues were observed for the JFM.

For the LWR, the magnitude of the average effect moved closer to the null when heterogeneity was higher and farther from the null when heterogeneity was low. Specifically, the mean WR was 1.214 under high heterogeneity (Scenario 3) versus 1.302 under low heterogeneity (Scenario 2). Relative bias was small ($\approx$0.36-1.90\%) across all scenarios, with the largest upward bias under high heterogeneity (Scenario 3, $+1.89\%$). Stratification on $Z_2$ had a negligible impact on bias or variability; stratified and unstratified analyses yielded nearly identical means and standard errors in every scenario. Across all scenarios, the unstratified analyses generated roughly 40,000 pairwise comparisons on average, whereas the stratified analyses yielded about 20,000. Ties were negligible in nearly all settings (typically $<\!1-3\%$), with the highest tie rate observed under the misspecification setting (Scenario 6), where the mean reached 11.39\%. Unlike the JFM, the LWR exhibited no convergence issues across the 3,000 simulated datasets, as it is a nonparametric method that does not involve model fitting. No runtime issues were observed for the LWR either.

Empirical power favored the JFM in all scenarios, including under baseline-hazard misspecification. JFM power ranged from 34\% (Scenario 4) to 98\% (Scenario 2), consistently exceeding the LWR by wide margins; for example, in the base scenario (Scenario 1), power was 82\% for the JFM versus 40.6\% for the LWR, and under misspecification, 56\% versus 40.4\%. The only instance where the LWR nominally matched or slightly exceeded the JFM was Scenario 4, by about 0.2 percentage points. Stratifying the LWR on $Z_2$ did not improve detection: power differences between stratified and unstratified versions were less than 0.4 percentage points across scenarios. Also, the highest power for the LWR was observed in Scenario 5 (70.6\%; high non-fatal event/low death rate scenario), where the mean WR was also the furthest from the null (1.356 vs. 1.207--1.302 in other scenarios). In this scenario, the mean number of recurrent events is the highest (3.73 per subject), twice as much as in the base scenario.

Because the empirical power results for the stratified LWR were unexpected, we evaluated an additional scenario in which data were generated with distinct baseline hazards across $Z_2$ strata. The settings were identical to Scenario 5, except that the baseline hazard rates were decreased by 0.1 when $Z_2 = 0$ and increased by 0.1 when $Z_2 = 1$. Under these conditions, compared with Scenario 5, the proportion of deaths slightly decreased (27.0\% vs. 29.5\%), while the mean number of recurrent events per subject slightly increased (3.8 vs. 3.7). The unstratified LWR achieved an empirical power of 73.8\%, whereas the stratified LWR reached 77.6\%.

Finally, we report $1/\mathrm{WR}$ alongside WR because some authors interpret the standard win ratio under proportional hazards as approximating a hazard ratio. In our simulations, the mean $1/\mathrm{WR}$ lay between 0.749 and 0.839 across scenarios, a range bracketed by the prespecified hazard ratios for the recurrent-event covariate $\exp(\beta^{R}_1)\approx0.70$, the terminal event $\exp(\beta^{D})\approx0.80$, and the second recurrent covariate $\exp(\beta^{R}_2)\approx0.90$ (Table \ref{tab:jfm_all}). Patterns were coherent with event composition: when recurrent events predominated (Scenario 5), $1/\mathrm{WR}$ moved toward 0.70 ($\mathrm{WR} = 0.749$); when deaths were frequent and truncated follow-up (Scenario 4), $1/\mathrm{WR}$ shifted toward 0.80 ($\mathrm{WR} = 0.839$). These comparisons are descriptive and included for context rather than as a formal calibration of WR to a hazard-ratio scale.

\section{Applications}\label{sec4}

\subsection{Studies}

To illustrate the methods presented in this article, we considered two clinical applications. The first study, available in the \texttt{frailtypack} package \cite{Krol2017-gs} and referred to as "Readmission", was a prospective cohort study meant to investigate sex differences in colorectal cancer-related hospital readmissions \cite{gonzalez_sex_2005}. The primary endpoint was the readmission time after colorectal cancer surgery. A total of 403 patients were recruited between 1996 and 1998 and randomized between the two treatment arms (new chemotherapy and control). The follow-up was until 2002, with a median follow-up of approximately 3 years. Using a Cox frailty model and adjusting for several covariates, the authors reported no significant effect of chemotherapy on the risk of readmission (HR = 1.34, 95\% CI 0.96 - 1.86) \cite{gonzalez_sex_2005}. The covariates included in the dataset were sex, the Charlson comorbidity index (CCI), and the Dukes classification. The Dukes classification describes the extent of colorectal cancer progression from localized (A) to metastatic (D) stages. The CCI is a prognostic measure that quantifies a patient's burden of comorbid conditions and predicts mortality risk. It ranges from 0 (no comorbidities) to 6 (high burden of comorbidities). In our dataset, it was treated as a time-dependent covariate and categorized into three levels: 0, 1-2, and $\ge$3.

The second study, available in the \texttt{WR}\cite{WR_package-rt} package referred to as "HF-ACTION", is a randomized controlled trial conducted on a cohort of over two thousand ($n = 2331$) heart failure patients recruited between 2003 -- 2007 for a median follow-up of 30 months \cite{OConnor2009-od}. The study aimed to assess the effect of adding aerobic exercise training to usual care on patient outcome. The primary endpoint was a composite measure that included both hospitalization (any cause) and death (any cause). The primary analysis demonstrated a moderate but non-significant reduction in the risk of time to the first composite event (hazard ratio 0.93; p-value 0.13). Here we focus on a subgroup of non-ischemic patients with reduced cardiopulmonary exercise (CPX) test duration (i.e., $\le$9 minutes before reporting discomfort). There is scientific and empirical evidence suggesting that this particular sub-population may derive greater benefit from exercise training interventions compared to the average heart failure patient \cite{rprojectTwosampleRatio}.

\subsection{Treatment effect}

For the JFM, we estimated baseline hazard functions semi-parametrically using percentile splines with six knots. We computed confidence intervals and p-values using the robust variance estimator.

In total, for the applications, estimations using JFMs were performed in approximately 6 minutes (3 minutes seconds for HF-ACTION, 3 minutes for Readmission) while the LWR analyses were completed in less than 1 second.

\subsubsection{Readmission dataset}

Among 403 patients (control, $n = 186$; experimental, $n = 217$), the median follow-up was 3.1 years for a maximum of approximately 6 years. The number of readmissions per patient ranged from 0 to 22, with a mean of 1.14 readmission per subject. Half of the patients experienced at least one readmission. Overall, 109 patients (27.0\%) died during follow-up; among them, 73 (67.0\%) had at least one readmission prior to death. Among the 294 survivors, 131 (44.6\%) experienced at least one readmission. In total, 204 of 403 patients (50.6\%) had at least one readmission.

\begin{table}[ht!]
  \small
  \centering
  \caption{Joint frailty model and win ratio analyses results on the Readmission dataset}
  \label{tab:readmission}

  \textbf{Panel A. Joint frailty model}\\[2pt]

  \renewcommand{\arraystretch}{1.2}
  \begin{tabular}{lrrrr}
    \hline
                                    & \multicolumn{2}{c}{\textbf{Recurrent events}} & \multicolumn{2}{c}{\textbf{Terminal event (death)}}                                                                                \\
    \cline{2-3}\cline{4-5}
    \textbf{Covariate / contrast}   & \multicolumn{1}{c}{\textbf{HR (95\% CI)}}     & \multicolumn{1}{c}{\textbf{$p$}}                    & \multicolumn{1}{c}{\textbf{HR (95\% CI)}} & \multicolumn{1}{c}{\textbf{$p$}} \\
    \hline
    \multicolumn{5}{l}{\emph{Unadjusted model} \quad--\quad $\theta=1.38$ ($p<0.001$), $\alpha = 1.28$ ($p < 0.001$)}                                                                                                    \\
    \quad Chemotherapy (yes vs. no) & 0.78 (0.52, 1.17)                             & 0.228                                               & 1.61 (0.90, 2.89)                         & 0.109                            \\[2pt]
    \multicolumn{5}{l}{\emph{Adjusted model} \quad--\quad  $\theta=1.07$ ($p<0.001$), $\alpha = 0.63$ ($p < 0.001$)}                                                                                                     \\
    \quad Chemotherapy (yes vs. no) & 0.86 (0.61, 1.20)                             & 0.371                                               & \phantom{2}2.90 (1.62, \phantom{7}5.18)   & <0.001                           \\
    \quad Sex (female vs. male)     & 0.53 (0.38, 0.73)                             & <0.001                                              & \phantom{2}0.78 (0.48, \phantom{7}1.28)   & 0.323                            \\
    \quad Charlson 1--2 vs. 0       & 1.63 (0.70, 3.79)                             & 0.257                                               & \phantom{2}1.62 (0.51, \phantom{7}5.20)   & 0.417                            \\
    \quad Charlson $\geq$3 vs. 0    & 1.79 (1.24, 2.59)                             & 0.002                                               & \phantom{2}3.60 (2.01, \phantom{7}6.47)   & <0.001                           \\
    \quad Dukes C vs. A/B           & 1.43 (0.95, 2.16)                             & 0.088                                               & \phantom{2}3.60 (1.61, \phantom{7}8.02)   & 0.002                            \\
    \quad Dukes D vs. A/B           & 4.67 (2.82, 7.72)                             & <0.001                                              & 21.90 (9.07,           52.91)             & <0.001                           \\
    \hline
  \end{tabular}

  \begin{tablenotes}
    \item \textbf{Notes.} Adjusted global tests: Charlson $p<0.001$; Dukes $p<0.001$. CIs are from $\exp(\beta \pm 1.96\,\mathrm{SE(HIH)})$.
  \end{tablenotes}
  \vspace{1em}
  \textbf{Panel B. Last-event-assisted win ratio} \\[2pt]

  \renewcommand{\arraystretch}{1.2}
  \makebox[\textwidth][c]{%
    \begin{tabular}{lrrrrr}
      \hline
      \textbf{Analysis}                     & \multicolumn{1}{c}{\textbf{\# pairs}} & \multicolumn{1}{c}{\textbf{Win prob. (\%)}} & \multicolumn{1}{c}{\textbf{Loss prob. (\%)}} & \multicolumn{1}{c}{\textbf{WR (95\% CI)}} & \multicolumn{1}{c}{\textbf{$p$}} \\
      \hline
      Unstratified                          & 40,362                                & 38.5                                        & 39.3                                         & 0.98 (0.75, 1.27)                         & 0.878                            \\
      Stratified by sex                     & 20,694                                & 38.8                                        & 39.4                                         & 0.98 (0.76, 1.28)                         & 0.897                            \\
      Stratified by Charlson                & 22,023                                & 34.3                                        & 43.4                                         & 0.79 (0.61, 1.03)                         & 0.076                            \\
      Stratified by Dukes                   & 12,519                                & 32.6                                        & 43.1                                         & 0.76 (0.57, 1.01)                         & 0.058                            \\
      Stratified by Dukes $\times$ Charlson & 7,975                                 & 33.2                                        & 42.4                                         & 0.78 (0.58, 1.05)                         & 0.104                            \\
      \hline
    \end{tabular}
  }
  \vspace{4pt}
  \begin{tablenotes}
    \item \textbf{Counts (control vs. experimental).} $N=186$ vs. $217$; recurrent events $=282$ vs. $176$; deaths $=51$ vs. $58$; median follow-up $=1258.5$ vs. $1054.0$ days.
    \item \textbf{Abbreviations.} HR, hazard ratio; WR, win ratio; CI, confidence interval; \#, number of.; prob., probability.
  \end{tablenotes}
\end{table}

Using the JFM with spline-based baseline hazards and robust variance (Table \ref{tab:readmission}, Panel A), the unadjusted analysis indicated substantial between-patient heterogeneity (frailty variance $\theta=1.38$, $p<0.001$) and a positive association between recurrent readmissions and death ($\alpha=1.28$, $p<0.001$). Chemotherapy showed opposite directions across endpoints: a non-significant reduction in the readmission rate ($\mathrm{HR} = 0.78$, 95\% CI = 0.52-1.17; $p=0.228$) and a non-significant increase in the hazard of death ($\mathrm{HR} = 1.61$, 95\% CI = 0.90 - 2.89; $p=0.109$). After adjustment for sex, Charlson index, and Dukes stage, both heterogeneity and association remained statistically significant but attenuated ($\theta=1.07$, $p<0.001$; $\alpha=0.63$, $p<0.001$), as expected. In the adjusted model, chemotherapy no longer influenced readmissions ($\mathrm{HR} = 0.86$, 95\% CI = 0.61-1.20; $p=0.371$) but was associated with a higher hazard of death ($\mathrm{HR} = 2.90$, 95\% CI = 1.62 - 5.18; $p<0.001$).

As for the LWR analysis (Table \ref{tab:readmission}, Panel B), across the unstratified analysis and all stratified variants (by sex, Charlson, Dukes, and their interaction), the WR estimates were below 1 but not statistically significant (e.g., unstratified $\mathrm{WR} = 0.98$, 95\% CI = 0.75 - 1.27; $p=0.878$). Stratification by baseline prognostic factors yielded similar conclusions (despite an increased magnitude of effect), with several estimates near but still above the two-sided 5\% threshold (e.g., $\mathrm{WR} = 0.76$, 95\% CI = 0.57 - 1.01 when stratified by Dukes; $p=0.058$). Note that, for the stratified analyses using the Charlson index, as it is a time-dependent covariate, we paired individuals based on their Charlson scores at baseline. Win ratio conclusions were consistent across the choice of win function (SWR, NWR, FWR, and LWR; Appendix Table \ref{tab:winfunc_sensitivity}). Only the NWR with stratification on Dukes stage yielded a significant win ratio estimate (p=0.049), still at the border of significance.

Overall, LWR results suggest no detectable net benefit of chemotherapy once both components and prioritization rules are taken into account, which is inconsistent with the JFM findings. These results could be explained by the opposing effects of chemotherapy on the two endpoints observed in the JFM analysis. The LWR method does not allow us to disentangle the effects on each component separately, making it difficult to interpret the results in the presence of such opposing effects. However, the direction of the effect ($\mathrm{LWR}<1$) is consistent across methods, suggesting a potential harmful effect of chemotherapy when accounting for both readmissions and death.

\subsubsection{HF-Action dataset}

Among 424 patients (control $n = 221$; experimental $n = 203$), the median follow-up was 28 months for a maximum of 53 months. The number of rehospitalizations per patient ranged from 0 to 26, with a mean of 2.4 rehospitalizations per subject, and half of the subjects experienced at least one rehospitalization. A total of 93 patients (22.0\%) died during follow-up; of these, 82 (88.2\%) experienced at least one rehospitalization. Among survivors ($n = 331$), 233 (70.4\%) experienced at least one rehospitalization. Overall, 315 (74.3\%) experienced at least one rehospitalization.

\begin{table}[ht!]
  \small
  \centering
  \caption{Joint frailty model and win ratio analyses results on the HF-ACTION dataset}
  \label{tab:hfaction}

  \textbf{Panel A. Joint frailty model}\\[2pt]

  \renewcommand{\arraystretch}{1.2}
  \begin{tabular}{lrrrr}
    \hline
                                  & \multicolumn{2}{c}{\textbf{Recurrent events}} & \multicolumn{2}{c}{\textbf{Terminal event (death)}}                                                                                \\
    \cline{2-3}\cline{4-5}
    \textbf{Covariate / contrast} & \multicolumn{1}{c}{\textbf{HR (95\% CI)}}     & \multicolumn{1}{c}{\textbf{$p$}}                    & \multicolumn{1}{c}{\textbf{HR (95\% CI)}} & \multicolumn{1}{c}{\textbf{$p$}} \\
    \hline
    \multicolumn{5}{l}{\emph{Unadjusted model} \quad--\quad  $\theta=1.00$, $p<0.001$, $\alpha = 1.37$ ($p < 0.001$)}                                                                                                  \\
    \quad Treatment (1 vs. 0)     & 0.78 (0.60, 1.01)                             & 0.058                                               & 0.60 (0.36, 1.00)                         & 0.052                            \\[2pt]
    \multicolumn{5}{l}{\emph{Adjusted model} \quad--\quad  $\theta=0.98$ ($p<0.001$), $\alpha = 1.57$ ($p<0.001$)}                                                                                                     \\
    \quad Treatment (1 vs. 0)     & 0.75 (0.58, 0.97)                             & 0.030                                               & 0.59 (0.35, 1.01)                         & 0.056                            \\
    \quad Age $\geq$60 vs. $<$60  & 0.72 (0.56, 0.93)                             & 0.011                                               & 1.47 (0.86, 2.52)                         & 0.159                            \\
    \hline
  \end{tabular}

  \vspace{4pt}
  \begin{tablenotes}
    \item \textbf{Note.} CIs are from $\exp(\beta \pm 1.96\,\mathrm{SE(HIH)})$.
  \end{tablenotes}
  \vspace{4pt}

  \textbf{Panel B. Last-event-assisted win ratio} \\[2pt]

  \renewcommand{\arraystretch}{1.2}
  \makebox[\textwidth][c]{%
    \begin{tabular}{lrrrrr}
      \hline
      \textbf{Analysis}                    & \multicolumn{1}{c}{\textbf{\# pairs}} & \multicolumn{1}{c}{\textbf{Win prob. (\%)}} & \multicolumn{1}{c}{\textbf{Loss prob. (\%)}} & \multicolumn{1}{c}{\textbf{WR (95\% CI)}} & \multicolumn{1}{c}{\textbf{$p$}} \\
      \hline
      Unstratified                         & 45,305                                & 50.3                                        & 38.5                                         & 1.31 (1.04, 1.64)                         & 0.023                            \\
      Stratified by age ($\geq$60 vs. <60) & 23,239                                & 50.4                                        & 38.2                                         & 1.32 (1.05, 1.66)                         & 0.019                            \\
      \hline
    \end{tabular}
  }

  \vspace{4pt}
  \begin{tablenotes}
    \item \textbf{Counts (control vs. experimental).} $N=221$ vs. $205$; recurrent events $=571$ vs. $451$; deaths $=57$ vs. $36$; median follow-up $=28.62$ vs. $27.57$ months.
    \item \textbf{Abbreviations.} HR, hazard ratio; CI, confidence interval; WR, win ratio; \#, number of.; prob., probability.
  \end{tablenotes}
\end{table}

For the JFM (Table \ref{tab:hfaction}, Panel A), the unadjusted analysis revealed that both the frailty variance and the association parameter were statistically significant ($\theta=1.00$, $p<0.001$; $\alpha=1.37$, $p<0.001$), and the point estimates for treatment (exercise training vs. usual care), indicated a protective direction on both endpoints (readmissions $\mathrm{HR} = 0.78$, 95\% CI = 0.60 - 1.01; $p=0.058$; death $\mathrm{HR} = 0.60$, 95\% CI = 0.36 - 1.00; $p=0.052$), even if borderline non-significant. Note that by using non-robust standard errors, these unadjusted effects would cross the 5\% threshold (readmissions $p=0.036$, death $p=0.048$), whereas robust standard errors yield borderline values. After adjustment for age, the protective effect on readmissions reached statistical significance ($\mathrm{HR} = 0.75$, 95\% CI = 0.58 - 0.97; $p=0.030$), but not on death ($\mathrm{HR} = 0.59$, 95\% CI = 0.35 - 1.01; $p=0.056$). Heterogeneity and association remained strong and statistically significant in the adjusted model ($\theta=0.98$, $p<0.001$; $\alpha=1.57$, $p<0.001$), with $\theta$ essentially unchanged by age adjustment, indicating that age does not account for the between-patient variability captured by the frailty, even if it strengthens the association between rehospitalization and death ($\alpha=1.37$ to $\alpha=1.57$).

The WR analyses corroborated a beneficial treatment effect on the composite with last-event assistance (Table \ref{tab:hfaction}, Panel B). The unstratified WR was 1.31 (95\% CI 1.04 - 1.65; $p=0.023$), indicating a significantly higher probability of "wins" than "losses" under the prespecified hierarchy. Results were stable when stratifying by age ($\mathrm{WR} = 1.32$, 95\% CI = 1.05 - 1.66; $p=0.019$). In this scenario, both methods lead to the conclusion that the treatment effect favored exercise training. Win ratio conclusions were also consistent across the choice of win function (SWR, NWR, FWR, and LWR; Appendix Table \ref{tab:winfunc_sensitivity_stratified}).

\subsection{Sample size calculation}

We designed a two-arm randomized controlled trial with 1:1 allocation, a 5\% two-sided type I error rate, and a target power of 80\% or 90\%, based on HF-Action results. We assumed exponential baseline hazards (i.e., constant hazard) for both recurrent and terminal events.  We assumed a median follow-up of 2.5 years, with uniform accrual over 3 years and an additional year of observation, yielding a maximum follow-up of 4 years.

\subsubsection{Schoenfeld's formula}

For the first sample size calculation, we relied on the Schoenfeld formula \cite{schoenfeld_sample-size_1983}:
\begin{align*}
  \mathrm{D} = \frac{(z_{1-\alpha_1/2} + z_{1-\alpha_2})^2}{\log(\mathrm{HR})^2 \times \mathrm{p}(1-\mathrm{p})},
\end{align*}
where $\mathrm{D}$ is the required number of events, $z_{1-\alpha_1/2}$ and $z_{1-\alpha_2}$ are the quantiles of the standard normal distribution for a two-sided type I error rate $\alpha_1$ and a target power $1-\alpha_2$, respectively, $\mathrm{HR}$ is the hazard ratio between the two arms, and $\mathrm{p}$ is the proportion of subjects in one arm (here, $\mathrm{p}=0.5$). Adapting the information presented in their study design paper \cite{Whellan2007-hk}, we assumed a hazard ratio of about 0.85 for the composite endpoint of all-cause death or hospitalization, leading to a required number of events of 1156 or 1548 for 80\% and 90\% power, respectively. To obtain a sample size, we may divide the number of required events by the mean probability of observing an event of approximately 54\% over the follow-up period. Details on how this probability was computed are provided in appendix \ref{app:schoenfeld}.

\subsubsection{Joint frailty model}

For the second sample calculation based on JFMs, additional parameters were required. The number of recurrent events was assumed to follow a Poisson distribution with a mean of three events per subject. We specified hazard ratios of 0.8 for recurrent events and 0.9 for the terminal event. The population was assumed to be heterogeneous with respect to the risk of recurrent events ($\theta = 1$), and the recurrent and terminal processes were positively associated ($\alpha = 1$).

Because this approach relies on Monte Carlo simulations, four additional parameters were required: the median time to death under the null and the shape parameter of the Weibull baseline hazard function, both for death and hospitalization. We assumed a median time to death of 28 months and a median time to first hospitalization of 3.6 months. As for the shape parameters, two scenarios were considered: scenario A with constant hazards over time (shapes = 1) and scenario B with increasing hazards (shapes > 1), with a faster increase for death (shape = 2) than for hospitalization (shape = 1.5). Accordingly, in the constant-hazard scenario A, the scale parameters were 40.40 and 5.19 for death and hospitalization, respectively; in the increasing-hazard scenario B, they were 33.63 and 4.60, respectively.

\subsubsection{Win ratio}

For the third sample size calculation, we followed the approach implemented by Mao et al. in their \texttt{WRSS} function, as illustrated in their package for the \texttt{WRSS} function \cite{WR_package-rt}. As this approach works for a two-component composite endpoint without recurrent events, we considered that only the first hospitalization was counted alongside death. First, we computed the Gumbel-Hougaard copula parameters using a subset of the HF-Action dataset composed of control patients. As before, we assumed a 3-year uniform accrual period and a maximum follow-up of 4 years. Because it is implemented in their package, an exponential loss-to-follow-up mechanism was considered, with a rate parameter of 0.05. The assumed hazard ratios were 0.9 for death and 0.8 for hospitalization. Based on these specifications, we derived the corresponding parameters required to apply the sample size formula \eqref{eq:ss_wrrec}.

However, it is not satisfactory to consider only the first hospitalization when recurrent events are of interest. For illustration purposes, we conducted a simulation-based approach using the LWR to determine an adequate sample size that accounts for recurrent hospitalizations. Data were generated under a similar framework as in the \texttt{frailtyDesign} family of functions of \texttt{frailtypack} \cite{Krol2017-gs}. Through simulations, we computed the empirical power (under the alternative) and type I error (under the null, i.e. both hazard ratios set to 1) over a grid of total sample sizes. The smallest sample size achieving at least 80\% power while maintaining the type I error at or below 5\% was selected. We compared two different scenarios: scenario A with the same specifications as before (frailty variance $\theta = 1$, hazard ratios of 0.8 for recurrent events and 0.9 for the terminal event., etc.) and scenario C with a lower frailty variance ($\theta = 0.5$) and higher treatment effects (hazard ratios of 0.7 for recurrent events and 0.8 for the terminal event.). As for the JFM-based sample size calculation, we did not consider an exponential loss-to-follow-up mechanism in the simulation-based approach, and instead inflated the computed sample sizes to account for a 5\% loss to follow-up.

\subsubsection{Computed sample sizes}

For the Schoenfeld and JFM-based computed sample sizes, lost to follow-up are not taken into account. Anticipating that 5\% of subjects would be lost to follow-up, we inflated the obtained sample sizes accordingly. We did not apply such an inflation for the WR-based sample sizes, as their implementation explicitly incorporates an exponential dropout mechanism. All final sample sizes were rounded up to the nearest even integer to ensure a balanced 1:1 allocation ratio; these results are summarized in Table \ref{tab:samplesize}. The code to compute the "standard win ratio" sample size ran for approximately 30 seconds; the JFM-based approach ran for approximately 9 minutes; and the simulation-based approach for LWR ran for approximately 4 minutes in scenario C, and more than 36 minutes in scenario A.
\begin{table}[h!]
  \centering
  \caption{Required total number of subjects by method and target power.}
  \label{tab:samplesize}
  \setlength{\tabcolsep}{8pt}
  \begin{tabular}{lrr}
    \toprule
                                   & \multicolumn{2}{c}{Sample size}              \\
    \cmidrule(lr){2-3}
    Method                         & 80\% power                      & 90\% power \\
    \midrule
    Schoenfeld                     & 2132                            & 2856       \\
    Joint frailty model            &                                 &            \\
    \hspace{1em}Constant hazards   & 1116                            & 1466       \\
    \hspace{1em}Increasing hazards & 832                             & 1092       \\
    Standard win ratio             & 1272                            & 1702       \\
    \bottomrule
  \end{tabular}
\end{table}

As expected, the Schoenfeld formula yielded the largest sample size requirements. This is due to the method's reliance on time-to-first-event data, which limits statistical power compared to approaches utilizing the full longitudinal history. Within the JFM framework, the increasing hazards scenario required fewer subjects than the constant hazards scenario; the accumulation of a higher number of events during follow-up under increasing hazards naturally enhances statistical power. However, it is important to acknowledge that JFM computations are sensitive to the selection of the frailty variance-a parameter subject to inherent uncertainty that can high influence sample size estimation \cite{dinart_sample_2024}. The WR-based approach yielded estimates lower than the Schoenfeld formula (as it leverages more information than time-to-first-event by incorporating information from both components of the CE) but higher than the JFM (as it does not account for recurrent events).

Under the simulation-based framework for the WR, the required sample size for 80\% power exhibited significant sensitivity to the underlying data generation scenario. After inflating for a 5\% loss to follow-up, the computed sample size exceeded 6000 subjects under scenario A ($\theta = 1$ and hazard ratios of 0.8 for recurrent events and 0.9 for the terminal event) but decreased to approximately 1260 subjects under scenario C ($\theta = 0.5$ and hazard ratios of 0.7 and 0.8, respectively). This high disparity highlights the impact of treatment effect size and heterogeneity on power calculations. Furthermore, the divergence of these results from the standard win ratio at 80\% power suggests that equation \ref{eq:ss_wrrec} is clearly insufficient in this context, as it accounts for neither unobserved heterogeneity nor the frequency of recurrent events. For comparison, the JFM-based approach estimated a requirement of only 288 subjects to achieve 80\% power under the second scenario C.

\section{Discussion}\label{sec5}

Composite endpoints that include recurrent non-fatal events alongside a terminal event create a semi-competing risks structure, in which both the frequency and timing of events carry important clinical relevance. In this work, we highlighted three main challenges in the analysis of such endpoints: (i) the need to leverage all available information by accounting for both recurrent events and death, (ii) within-subject dependence due to unobserved heterogeneity and state dependence and (iii) the informative nature of death with respect to the recurrent events. We compared two frameworks that seemed promising for such outcomes: the recurrent-event WR (especially, the LWR) and the JFM. In particular, JFM directly addresses all those challenges by construction. Across six simulation scenarios and two applications, both approaches were empirically well-calibrated; however, JFM exhibited substancially higher power than LWR under our data-generating processes.

From a computational perspective, JFM estimation relies on numerical integration and optimization, whereas the win ratio requires forming all treated-control pairs. In our simulations and applications, both approaches ran without runtime issues; nevertheless, JFM estimation was always way slower than the win ratio. Same conclusion holds for the sample size calculations.

In terms of interpretation, under unmatched pairing in a randomized clinical trial (RCT), the win proportion consistently estimates the population probability that a randomly drawn treated patient "wins" against a randomly drawn control; i.e., a population-level estimand. For the JFM, under randomization (or no unmeasured confounding), correct model specification, and independent censoring, the regression vectors $\beta_R, \beta_D$ have a subject-specific causal interpretation as log hazard ratios conditional on the frailty $\omega_i$ and covariates; they are not marginal hazard ratios because frailty induces non-PH at the population level.

JFM essentially yielded unbiased subject-specific estimates for the component-specific log-hazards (recurrent and terminal), conditional on the subjects' frailty, while achieving high statistical power (see Table \ref{tab:jfm_all}). The frailty variance parameters remained the most difficult parameters to estimate in each scenario. Note that we modeled the baseline risk parametrically using a Weibull distribution. This approach was convenient for exploring the case of model misspecification (Scenario 6). However, except under stronger assumptions, semi-parametric modeling through splines is generally preferable, as done in our applications to clinical datasets. The LWR, in contrast, delivered a single population-level measure. We observed an attenuation of the LWR toward 1 as unobserved heterogeneity increased (scenarios 1, 2 and 3; see Table \ref{tab:lwr_all}). Likewise, power decreased with increasing frailty variance; this accords with Mao et al.'s simulations \cite{Mao2022-qd} showing reduced power for LWR as frailty variance increases. Likewise, we observed a strengthening of the estimated LWR for a high enough number of recurrent events (Scenario 5) and an attenuation when few recurrent events were observed (Scenario 4).

Even if both methods correctly identify the treatment effect, JFM outperformed LWR in terms of statistical power in most of our scenarios (with Scenario 4 an exception where both methods had similar power). In our simulation settings, this implies that, for a fixed sample size, JFM was more likely to detect a significant treatment effect than LWR. The LWR performed the best in scenario 5, where a mean of 3.73 recurrent events per subject was observed, twice as many as in the first scenario. Also, no tangible gains were observed when using stratified LWR over unstratified LWR, unlike in Mao et al. \cite{Mao2022-qd}. This pattern most likely arises from the data generation process, where $Z_2$ does not influence the baseline hazards. We confirmed this with the additional scenario, in which the baseline hazard was made dependent on $Z_2$. Therefore, stratification on a covariate should improve the power of the LWR when that covariate modifies the baseline hazard, but not otherwise.

The limited power of the LWR likely stems from insufficient separation between the distributions of wins and losses across the follow-up period. This overlap may result from three concurrent factors: a high death rate (except in Scenario 5 of "low death rate", where it was below 30\%), unobserved heterogeneity induced by the frailty component, and a relatively small sample size that prevents substantial reduction of the estimated standard error. Together, these factors may blur the contrast between the two treatment groups, thereby preventing a clear separation between the distribution of "wins" and "losses".

\subsection{Practical guidelines}

Taken together, we suggest the following strategy, emphasizing that our power comparisons are based on a limited set of simulation scenarios and should be treated as illustrative rather than universal. We recall that we consider composite endpoints of recurrent events and death.
\begin{itemize}
  \item \textbf{Assess available information.} To use either the WR or JFMs, researchers must have the specific event times for both recurrent events and death. If only the time to the first event is recorded, analyses are restricted to standard survival methods.
  \item \textbf{For WR analyses.} First, establish a clinical hierarchy, typically prioritizing death over recurrent events. Next, select a comparison strategy based on baseline patient information. An unmatched (all-pairs) approach is the standard default for randomized trials. Stratified matching is useful when specific baseline factors strongly influence risk, while one-to-one matched pairing (e.g., via risk scores) can be used but may exclude some patients. After choosing a matching strategy, select a win function (e.g., SWR, NWR, FWR, or LWR) that reflects the clinical definition of patient benefit (for instance, FWR emphasizes early events). We strongly recommend performing sensitivity analyses using different win functions and clearly reporting the total number of pairs along with the win, loss, and tie fractions. In this work, simulations and applications focus on unmatched and stratified pairing with LWR. Remember that WR estimands depend on both the win rule and the censoring/follow-up scheme and are therefore not censoring-robust in the sense of ICH E9(R1) \cite{Mao2022-qd}.
  \item \textbf{For JFM analyses.} Clearly state the chosen time scale (calendar time or gap time), how the baseline hazard is modeled (e.g., parametric or splines), and whether robust standard errors are used. When planning study size, follow the principles from Dinart et al. \cite{dinart_sample_2024}: (i) for a fixed total count of recurrent events, more participants with fewer events each yields higher power than fewer participants with more events; (ii) shortening follow-up while recruiting more participants is preferable to extending follow-up to "chase" extra recurrences; (iii) greater unobserved heterogeneity inflates the required sample size -- minimize it where possible; and (iv) the association parameter $\alpha$ between recurrent and terminal events should not be assigned too much importance at the design stage. This is because $\alpha$ is difficult to calibrate and mainly contributes to model interpretation. Therefore, several values of $\alpha$ should be explored in sensitivity analyses according to the expected direction of association: $\alpha=0$ if no clinical association between recurrence and death is expected, $\alpha>0$ if a higher recurrence risk is expected to be associated with a higher death risk, and $\alpha<0$ if the opposite is expected. A default value such as $\alpha=1$ may be used when a positive association is expected. Because the association parameter $\alpha$ is defined as a power of the frailties, it is not meaningful when the frailty variance is zero.

  \item \textbf{Planning and power considerations.} In our simulations, JFM generally achieved higher statistical power than the last-event win ratio (LWR), especially when recurrent events were moderately frequent. However, the best method depends on specific trial characteristics, such as event rates, follow-up length, patient heterogeneity, and censoring patterns. When designing a study, we recommend running simulations across a range of realistic scenarios to select the most suitable approach.
\end{itemize}
Finally, we recommend JFM over WR as the primary analysis if the treatment might have opposing effects on different outcomes (e.g., reducing recurrences but increasing mortality). JFM is also the better choice if researchers need to estimate the treatment effect on each component separately, adjust for confounding variables, or explore how specific covariates modify the treatment effect, as the WR framework does not easily support these detailed breakdowns.

\subsection{Limitations and future work}

Table \ref{tab:jfm-vs-wr} below summarizes the strengths and limitations of both approaches. For the sample size calculation, it is not (yet) possible to compute a sample size for recurrent events using Mao's approach. Future works could aim to develop methods for this purpose, but for now, we proposed a simulation-based approach.
For a standard win ratio, the framework of Mao et al. \cite{Mao2022-zj} (model-based approach) may not be convenient, as it requires parameters that are not easily interpretable clinically nor obtainable without prior data. A similar downside exists with the JFM-based approach, as it requires Monte-Carlo simulations \cite{dinart_sample_2024}, hence the need to specify parameters underlying the data-generating process -- notably for the gamma-frailty variance and the strength of the association between recurrent and terminal events. In practice, we therefore recommend sensitivity analyses that vary key inputs (e.g., the frailty variance for JFM) over plausible ranges. An additional, practical distinction concerns censoring: in the WR framework of Mao, loss to follow-up is explicitly taken into account through an exponential censoring model when computing sample size. By contrast, the current sample size function for JFMs in \texttt{frailtypack} does not directly parameterize censoring -- hence our 5\% inflation in the computed sample size.

The main limitations of JFM lie in its reliance on specific distributional assumptions -- such as the gamma frailty and the parameterization of the baseline hazard -- as well as the lack of a mechanism to incorporate clinical hierarchy among events. Also, too few recurrent events may lead to numerical issues in estimation. On the other hand, the recurrent-event win-ratio estimands depend on the censoring distribution; as such, they are best treated as hypothesis-testing tools. To account for the censoring in the standard win ratio, IPCW (Inverse Probability Censoring Weighting) variants \cite{Dong2020-eg} have been developed. That consists of weighting each treated-control pair win/loss outcome with the inverse probability that both patients are uncensored up to the pair-specific comparison time. Providing an IPCW-adjusted variant is a promising direction for this recurrent event win ratio. As it was done for the standard win ratio, extensions like recurrent-event win odds \cite{Brunner2021-bt} (which consists of distributing half of the ties to the numerator and the other half to the denominator) or net benefit \cite{Buyse2010-dc} could be developed to provide complementary measures of effect. Notably, for the recurrent-event net benefit, one could have an absolute measure (instead of a relative one as in the WR or win odds) that also allows for a decomposition of the effect on each component of the composite endpoint (i.e., recurrent events and death). Also, for win-ratio-type statistics, a recent work by Josse \& Even \cite{Even2025-ix} formalized hierarchical comparison within a counterfactual framework and clarified that the estimand depends on how pairs are formed. Using the unmatched pairing (what they call "complete" pairing) in an RCT, the classical win proportion consistently estimates a population-level probability that a randomly drawn treated patient wins over a randomly drawn control. It is an estimand that may depart from the ideal but non-identifiable individual-level probability that a given patient would fare better if treated than if untreated. In heterogeneous populations this mismatch can even reverse qualitative conclusions \cite{Even2025-ix}. Josse \& Even therefore proposed an identifiable individual-level causal estimand and showed that using nearest-neighbor stratification to construct pairs in an RCT yielded a consistent estimator of this estimand. Their framework also extends to observational settings via propensity weighting and distributional regression, with double-robust variants. In future works, it could be interesting to extend the work by Mao et al. \cite{Mao2022-qd} to accommodate IPCW and target an identifiable individual-level causal estimand, using Josse \& Even's nearest-neighbor matching approach \cite{Even2025-ix}, and to compare the resulting estimands and estimators with those from JFMs.

\begin{table}[ht!]
  \caption{Strengths and limitations of joint frailty models (JFM) vs. win ratio (WR) for composite endpoints with recurrent events}
  \label{tab:jfm-vs-wr}
  \centering
  \makebox[\textwidth][c]{%
    \begin{tabular}{p{0.11\textwidth} p{0.42\textwidth} p{0.42\textwidth}}
      \hline
      \textbf{Category}    & \textbf{Joint frailty model (JFM)} & \textbf{Win ratio for recurrent events (WR)} \\
      \hline
      \vspace{0.1em}
      \textbf{Strengths}   &
      \begin{itemize} 
        \item Flexible modeling framework
        \item Decomposes effects on recurrent events vs. death
        \item Models unobserved heterogeneity ($\theta$)
        \item Accounts for dependence between processes ($\theta$, $\alpha$)
        \item Naturally handles censoring and truncation
        \item Accommodates (time-dependent, or not) covariate adjustment and interactions
        \item Enables dynamic prediction
        \item Available software: \texttt{frailtypack} \cite{Krol2017-gs}
      \end{itemize}
                           &
      \begin{itemize} 
        \item Intuitive interpretation
        \item Easy to implement
        \item Respects prespecified clinical hierarchy
        \item Requires no modeling assumptions (e.g., no proportional hazard assumption)
        \item Available software: \texttt{WR} \cite{WR_package-rt}
      \end{itemize}
      \\ \\
      \vspace{0.1em}
      \textbf{Limitations} &
      \begin{itemize} 
        \item Requires modeling assumptions; risk of misspecification (baseline risk, frailty distribution)
        \item Estimation can be computationally intensive
        \item Does not account for the clinical hierarchy
        \item Needs sufficient number of recurrent events per subject for stable estimation
        \item Sample size calculation relies on parameters that may be difficult to clinically interpret
      \end{itemize}
                           &
      \begin{itemize} 
        \item Requires a prespecified and clinically relevant component hierarchy
        \item Provides a global summary measure; cannot separate effects for recurrences vs. death
        \item For lower sample sizes, a strong size effect seems needed to detect a significant effect on the composite
        \item Limited covariate adjustment options (stratification only)
        \item Sample size calculation is not implemented for recurrent-events
        \item Even for the standard WR, sample size calculation remains difficult to perform without prior data (for the required parameters)
        \item Methodological development and software support remain yet immature
      \end{itemize}
      \\
      \hline
    \end{tabular}
  }
\end{table}

\section*{Data and code availability} \label{sec6}

The code used to generate the results of this article is publicly available at the following GitHub repository: \url{https://github.com/adrienorue/comparison_wrRec_jfm}. The repository contains R scripts to reproduce the simulation study, to perform the analyses of the two clinical applications, and to conduct the sample size calculations. The modified version of Mao et al.'s \texttt{WR} package is also available, which now includes the simulation-based sample size computation using the LWR (\texttt{sz\_lwr()} function). The code is licensed under the MIT license. The application datasets used in this study are available in the two packages \texttt{frailtypack} \cite{Krol2017-gs} (for Readmission) and \texttt{WR} \cite{WR_package-rt} (for HF-ACTION). To facilitate reproducibility, we also provide a master script that centralizes the workflow needed to reproduce the manuscript results, from the simulation study to the applications. By default, the script reproduces Scenario 1; other scenarios can be reproduced by modifying the documented scenario-specific inputs. The simulation-based LWR sample-size calculation is disabled by default because it is computationally intensive. Sample code for the analysis of the Readmission dataset and for sample size estimation is provided in Appendix \ref{app:r_code} and Appendix \ref{app:r_code_ss}, respectively. The master script is presented in Appendix \ref{app:master_script}, and is available in the GitHub repository.

\section*{Financial disclosure}

With the financial support of the French National Cancer Institute (Institut National du Cancer, INCa grant INCa$\_$18610).

\section*{Conflict of interest}

The authors declare no potential conflict of interests.

\newpage

\bibliographystyle{unsrtnat}
\bibliography{supp/references}

\section{Appendix}

\subsection{Unstratified win ratio variance}\label{app:wr_unstrat}

We recall that $\hat w$ and $\hat \ell$ are the observed fractions of wins and losses, respectively. By two-sample U-statistic theory and the delta method, one has the following asymptotic distribution:
\begin{align*}
  \sqrt{n}\,\Bigl[\log\widehat{\mathrm{WR}}_{\mathrm{rec}}(W)-\log\mathrm{WR}_{\mathrm{rec}}(W)\Bigr]\ \xrightarrow{d}\ \mathcal{N}\!\bigl(0,\sigma^2(W)\bigr)
\end{align*}
A consistent estimator of $\sigma^2(W)$ is
$$
  \hat\sigma^2(W)=\hat w^{-2}\,e_1^\top\hat\Sigma e_1
    - 2(\hat w\hat\ell)^{-1}\,e_1^\top\hat\Sigma e_2
  + \hat \ell^{-2}\,e_2^\top\hat\Sigma e_2,
$$
where $e_1=(1,0)^\top$, $e_2=(0,1)^\top$ and
$$
  \hat\Sigma=\frac{n}{n_{\mathcal{E}}^2}\sum_{e\in\mathcal{E}}\mathbf{u}_{t,e}\mathbf{u}_{t,e}^\top
  +\frac{n}{n_{\bar{\mathcal{E}}}^2}\sum_{\bar{e}\in\bar{\mathcal{E}}}\mathbf{u}_{c,\bar{e}}\mathbf{u}_{c,\bar{e}}^\top,
$$
with influence terms
$$
  \mathbf{u}_{t,e}=
  \begin{pmatrix}
    \frac{1}{n_{\bar{\mathcal{E}}}}\sum_{\bar{e}\in\bar{\mathcal{E}}} W\!\bigl(Y_e,Y_{\bar{e}};\tau_{e,\bar{e}}\bigr)-\hat w \\[0.25em]
    \frac{1}{n_{\bar{\mathcal{E}}}}\sum_{\bar{e}\in\bar{\mathcal{E}}} W\!\bigl(Y_{\bar{e}},Y_e;\tau_{e,\bar{e}}\bigr)-\hat \ell
  \end{pmatrix},
  \quad
  \mathbf{u}_{c,\bar{e}}=
  \begin{pmatrix}
    \frac{1}{n_{\mathcal{E}}}\sum_{e\in\mathcal{E}} W\!\bigl(Y_e,Y_{\bar{e}};\tau_{e,\bar{e}}\bigr)-\hat w \\[0.25em]
    \frac{1}{n_{\mathcal{E}}}\sum_{e\in\mathcal{E}} W\!\bigl(Y_{\bar{e}},Y_e;\tau_{e,\bar{e}}\bigr)-\hat \ell
  \end{pmatrix}.
$$

\subsection{Stratified win ratio variance}\label{app:wr_strat}

As in the unstratified case, we have the following asymptotic distribution:
$$
  \sqrt{n}\,\Bigl[\log\bigl(\widehat{\mathrm{WR}}_{\mathrm{str}}\bigr)-\log\bigl(\mathrm{WR}_{\mathrm{str}}\bigr)\Bigr]\ \xrightarrow{d}\ \mathcal{N}\!\bigl(0,\sigma_{\mathrm{str}}^2(W)\bigr).
$$
Analogously, a consistent estimator of $\sigma_{\mathrm{str}}^2(W)$ is
$$
  \hat\sigma^2_{\mathrm{str}}(W)=
  \hat w_{\mathrm{str}}^{-2}\,e_1^\top\hat\Sigma_{\mathrm{str}}e_1
    -2(\hat w_{\mathrm{str}}\hat \ell_{\mathrm{str}})^{-1}\,e_1^\top\hat\Sigma_{\mathrm{str}}e_2
  +\hat \ell_{\mathrm{str}}^{-2}\,e_2^\top\hat\Sigma_{\mathrm{str}}e_2,
$$
where
$$
  \hat\Sigma_{\mathrm{str}}
  = n\sum_{s=1}^S (\omega^{(s)})^2\!
  \left[
    \frac{1}{\bigl(n_{\mathcal{E}}^{(s)}\bigr)^2}\sum_{e=1}^{n_{\mathcal{E}}^{(s)}}\mathbf{u}^{(s)}_{t,e}\mathbf{u}^{(s)\top}_{t,e}
    +\frac{1}{\bigl(n_{\bar{\mathcal{E}}}^{(s)}\bigr)^2}\sum_{\bar{e}=1}^{n_{\bar{\mathcal{E}}}^{(s)}}\mathbf{u}^{(s)}_{c,\bar{e}}\mathbf{u}^{(s)\top}_{c,\bar{e}}
    \right],
$$
and
$$
  \mathbf{u}^{(s)}_{t,e}=
  \begin{pmatrix}
    \frac{1}{n_{\bar{\mathcal{E}}}^{(s)}}\sum_{\bar{e}} W\!\bigl(Y_e^{(s)},Y_{\bar{e}}^{(s)};\tau_{e,\bar{e}}^{(s)}\bigr)-\hat w^{(s)} \\[0.2em]
    \frac{1}{n_{\bar{\mathcal{E}}}^{(s)}}\sum_{\bar{e}} W\!\bigl(Y_{\bar{e}}^{(s)},Y_e^{(s)};\tau_{e,\bar{e}}^{(s)}\bigr)-\hat \ell^{(s)}
  \end{pmatrix},
  \quad
  \mathbf{u}^{(s)}_{c,\bar{e}}=
  \begin{pmatrix}
    \frac{1}{n_{\mathcal{E}}^{(s)}}\sum_{e} W\!\bigl(Y_e^{(s)},Y_{\bar{e}}^{(s)};\tau_{e,\bar{e}}^{(s)}\bigr)-\hat w^{(s)} \\[0.2em]
    \frac{1}{n_{\mathcal{E}}^{(s)}}\sum_{e} W\!\bigl(Y_{\bar{e}}^{(s)},Y_e^{(s)};\tau_{e,\bar{e}}^{(s)}\bigr)-\hat \ell^{(s)}
  \end{pmatrix}.
$$

\subsection{Pseudocode for dataset generation of the simulation study}\label{app:gen_proc}

\begin{algorithm}
  \caption{\enskip Simulation of recurrent and terminal events under a joint frailty model with exponentially distributed baseline hazards}\label{alg:simulation}
  \begin{algorithmic}[1]
    \State Set number of datasets $K = 500$, number of subjects $N = 400$, and fixed censoring time $C=3$
    \State Set parameters: $\theta$, $\alpha=1$, $\beta^{R}_1=\log(0.7)$, $\beta^{R}_2=\log(0.9)$, $\beta^{D} = \log(0.8)$, $r_0$, $\lambda_0$
    \For{$k = 1$ to $K$}
    \State Set seed to $k$ for reproducibility
    \For{$i = 1$ to $N$}
    \State Sample frailty $u_i \sim \text{Gamma}(shape = 1/\theta, scale = \theta)$
    \State Generate the two binary covariates $(Z_{1, i}, Z_{2,i}) \sim \text{Bernoulli}(0.5)$
    \State Compute $\eta^{\text{rec}}_i = \beta^{R}_1 Z_{1,i} + \beta^{R}_2 Z_{2,i} + \log(u_i)$
    \State Compute $\eta^{\text{death}}_i = \beta^{D} Z_{1,i} + \alpha \log(u_i)$
    \State Sample $T^{\text{death}}_i \sim \text{Exp}(\lambda_0 \cdot e^{\eta^{\text{death}}_i})$
    \State Initialize $t \gets 0$
    \While{$t < \min(T^{\text{death}}_i, C)$}
    \State Sample gap time $\Delta t \sim \text{Exp}(r_0 \cdot e^{\eta^{\text{rec}}_i})$
    \State Set $t_{\text{next}} \gets t + \Delta t$
    \If{$t_{\text{next}} < T^{\text{death}}_i$ and $t_{\text{next}} < C$}
    \State Record recurrent event at $t_{\text{next}}$
    \State $t \gets t_{\text{next}}$
    \Else
    \If{$T^{\text{death}}_i < C$}
    \State Record terminal event at $T^{\text{death}}_i$
    \Else
    \State Record censoring at $C$
    \EndIf
    \State \textbf{break}
    \EndIf
    \EndWhile
    \EndFor
    \EndFor
  \end{algorithmic}
\end{algorithm}

\subsection{Sensitivity to win-function choice}\label{app:winfunc_sensitivity}

To assess whether win-ratio conclusions were robust to the choice of the win function, we recomputed the win ratio in the two applications using four variants: the standard win ratio (SWR; based on time to first non-fatal event), the naive win ratio (NWR), the first-event-assisted win ratio (FWR), and the last-event-assisted win ratio (LWR). Table \ref{tab:winfunc_sensitivity} reports unstratified estimates, and Table \ref{tab:winfunc_sensitivity_stratified} reports stratified estimates. SWR is computed by keeping only the first non-fatal event for each subject, so that FWR and LWR reduce to the standard win ratio.

In the unstratified analyses, all win functions produced similar conclusions. In the stratified analyses, only the NWR stratified by Dukes stage yielded a statistically significant win ratio estimate (p = 0.049), although the result remained borderline significant. Overall, these results suggest that the conclusions of the win ratio analyses were robust to the choice of the win function.

\begin{table}[h!]
  \begingroup
  \centering
  \caption{Sensitivity of win ratio estimates to the win-function choice in the two applications (unstratified).}
  \label{tab:winfunc_sensitivity}
  \small
  \setlength{\tabcolsep}{6pt}
  \begin{tabular}{llccc}
    \toprule
    Dataset     & Win function                  & WR   & 95\% CI      & $p$-value \\
    \midrule
    Readmission & SWR (time to first non-fatal) & 0.98 & (0.75, 1.28) & 0.887     \\
    Readmission & NWR (count)                   & 0.97 & (0.74, 1.27) & 0.815     \\
    Readmission & FWR (first-event-assisted)    & 0.98 & (0.75, 1.27) & 0.870     \\
    Readmission & LWR (last-event-assisted)     & 0.98 & (0.75, 1.27) & 0.878     \\
    HF-ACTION   & SWR (time to first non-fatal) & 1.26 & (1.00, 1.60) & 0.049     \\
    HF-ACTION   & NWR (count)                   & 1.33 & (1.04, 1.70) & 0.023     \\
    HF-ACTION   & FWR (first-event-assisted)    & 1.30 & (1.04, 1.64) & 0.024     \\
    HF-ACTION   & LWR (last-event-assisted)     & 1.31 & (1.04, 1.64) & 0.023     \\
    \bottomrule
  \end{tabular}
  \begin{tablenotes}
    \item \textbf{Notes.} SWR is computed by restricting each subject to their first non-fatal event (so that FWR/LWR reduce to the classic win ratio).
  \end{tablenotes}
  \endgroup
\end{table}

\begin{table}[h!]
  \begingroup
  \centering
  \caption{Sensitivity of win ratio estimates to the win-function choice in the two applications (stratified).}
  \label{tab:winfunc_sensitivity_stratified}
  \small
  \setlength{\tabcolsep}{6pt}
  \begin{tabular}{lllccc}
    \toprule
    Dataset     & Stratification variable & Win function                  & WR   & 95\% CI      & $p$-value \\
    \midrule
    Readmission & Sex                     & SWR (time to first non-fatal) & 0.98 & (0.76, 1.28) & 0.903     \\
    Readmission & Sex                     & NWR (count)                   & 0.97 & (0.74, 1.28) & 0.850     \\
    Readmission & Sex                     & FWR (first-event-assisted)    & 0.98 & (0.76, 1.28) & 0.893     \\
    Readmission & Sex                     & LWR (last-event-assisted)     & 0.98 & (0.76, 1.28) & 0.897     \\
    Readmission & Charlson                & SWR (time to first non-fatal) & 0.79 & (0.61, 1.03) & 0.085     \\
    Readmission & Charlson                & NWR (count)                   & 0.78 & (0.59, 1.02) & 0.072     \\
    Readmission & Charlson                & FWR (first-event-assisted)    & 0.79 & (0.61, 1.02) & 0.075     \\
    Readmission & Charlson                & LWR (last-event-assisted)     & 0.79 & (0.61, 1.03) & 0.076     \\
    Readmission & Dukes                   & SWR (time to first non-fatal) & 0.76 & (0.57, 1.02) & 0.065     \\
    Readmission & Dukes                   & NWR (count)                   & 0.74 & (0.54, 1.00) & 0.049     \\
    Readmission & Dukes                   & FWR (first-event-assisted)    & 0.75 & (0.56, 1.01) & 0.056     \\
    Readmission & Dukes                   & LWR (last-event-assisted)     & 0.76 & (0.56, 1.01) & 0.058     \\
    Readmission & Dukes x Charlson        & SWR (time to first non-fatal) & 0.79 & (0.59, 1.06) & 0.120     \\
    Readmission & Dukes x Charlson        & NWR (count)                   & 0.77 & (0.56, 1.05) & 0.097     \\
    Readmission & Dukes x Charlson        & FWR (first-event-assisted)    & 0.78 & (0.58, 1.05) & 0.102     \\
    Readmission & Dukes x Charlson        & LWR (last-event-assisted)     & 0.78 & (0.58, 1.05) & 0.104     \\
    HF-ACTION   & Age                     & SWR (time to first non-fatal) & 1.27 & (1.00, 1.60) & 0.049     \\
    HF-ACTION   & Age                     & NWR (count)                   & 1.34 & (1.05, 1.72) & 0.019     \\
    HF-ACTION   & Age                     & FWR (first-event-assisted)    & 1.32 & (1.04, 1.66) & 0.020     \\
    HF-ACTION   & Age                     & LWR (last-event-assisted)     & 1.32 & (1.05, 1.66) & 0.019     \\
    \bottomrule
  \end{tabular}
  \begin{tablenotes}
    \item \textbf{Notes.} SWR is computed by restricting each subject to their first non-fatal event (so that FWR/LWR reduce to the classic win ratio).
  \end{tablenotes}
  \endgroup
\end{table}

\newpage

\subsection{R code sample used in Readmission analysis}\label{app:r_code}

This appendix provide a sample code meant to illustrate the main steps of the analyses performed in the Readmission application. To reproduce our results, we highly recommend to see Section \ref{sec6}.

\begin{lstlisting}[
  language=R, 
  caption={Sample R code for the analysis of the Readmission dataset}, 
  commentstyle=\ttfamily\upshape, 
  keywordstyle=\ttfamily\mdseries,
  basicstyle=\fontsize{8}{10}\selectfont\ttfamily
]
  
# ----------------------------
# Joint frailty models
# ----------------------------
library(frailtypack) #please install the latest version from CRAN
data(readmission)
reducedRecurrent <- frailtyPenal(
    Surv(t.start, t.stop, event) ~ cluster(id) + as.factor(chemo) + sex + 
                                   charlson + dukes,
    recurrentAG = TRUE, #recurrentAG because calendar (t.start, t.stop)
    hazard = "Splines-per", kappa = 1e5, n.knots = 6, #6 = quartiles
    nb.gl = 50, #Gauss-Laguerre nodes (50, 32, 20). 50 = better precision
    cross.validation = TRUE, #choose kappa by CV
    data = readmission
)

reducedTerminal <- frailtyPenal(
    Surv(t.stop, death) ~ as.factor(chemo) + sex + charlson + dukes,
    hazard = "Splines-per", kappa = 1e5, n.knots = 6,
    nb.gl = 50,
    cross.validation = TRUE,
    data = readmission[readmission$event == 0, ] #last observation = death or censoring
)

initBetas <- unname(c(reducedRecurrent$coef, reducedTerminal$coef))
initTheta <- reducedRecurrent$theta
initKappas <- c(reducedRecurrent$kappa, reducedTerminal$kappa)

fitJFM_adjust <- frailtyPenal(
    Surv(t.start, t.stop, event) ~ cluster(id) + as.factor(chemo) + sex + 
                                   charlson + dukes + terminal(death),
    formula.terminalEvent =~ as.factor(chemo) + sex + charlson + dukes,
    recurrentAG = TRUE,
    hazard = "Splines-per", n.knots = 6,
    kappa = initKappas, init.B = initBetas, init.Alpha = 1,
    nb.gl = 50,
    data = readmission
)

# ----------------------------
# Win ratio
# ----------------------------
library(WR) 
readmission$stratifVar <- interaction(readmission$dukes, readmission$charlson)
readmission$stratifVar <- as.numeric(readmission$stratifVar) - 1
WRrec(
    ID = as.numeric(readmission$id),
    time = as.numeric(readmission$t.stop),
    status = as.numeric(readmission$status),
    trt = as.numeric(readmission$chemo),
    naive = TRUE, # to compute NWR and FWR variants
    strata = as.numeric(readmission$stratifVar) #remove if unstratified
)
\end{lstlisting}

\subsection{R code for sample size estimation} \label{app:r_code_ss}

This appendix provide a sample code meant to illustrate the main steps performed for sample size computation. To reproduce our results, we highly recommend to see Section \ref{sec6}.

\begin{lstlisting}[
  language=R, 
  caption={Sample R code for sample size estimation}, 
  commentstyle=\ttfamily\upshape, 
  keywordstyle=\ttfamily\mdseries,
  basicstyle=\fontsize{8}{10}\selectfont\ttfamily
]

# "Schoenfeld" ----------------------------------------------------------------
nbEvent <- function(alpha, power, p, HR) {
  z_alpha <- qnorm(1 - alpha / 2)
  z_beta <- qnorm(power)
  d <- (z_alpha + z_beta)^2 / (log(HR)^2 * p * (1 - p))
  return(ceiling(d))
}

lambdaC <- -log(1 - 0.3)
lambdaT <- -1 / 2 * log(1 - 0.89 * (1 - exp(-2 * lambdaC)))
HRest <- lambdaT / lambdaC
nbEvent80P <- nbEvent(alpha = 0.05, power = 0.8, p = 0.5, HR = HRest)
nbEvent90P <- nbEvent(alpha = 0.05, power = 0.9, p = 0.5, HR = HRest)

probaEventTreat <- 1 - (exp(-lambdaT) - exp(-4 * lambdaT)) / (3 * lambdaT)
probaEventControl <- 1 - (exp(-lambdaC) - exp(-4 * lambdaC)) / (3 * lambdaC)
meanProbaEvent <- (probaEventTreat + probaEventControl) / 2

nbEvent80P / meanProbaEvent # round => 2132
nbEvent90P / meanProbaEvent #round + nearest odd => 2856


# Standard win ratio sample size ----------------------------------------------
library(WR)

dat <- read.csv("hfaction.csv", sep = ",") # please replace with the correct path
pilot <- dat[dat$trt_ab == 0, ] # control group
id <- pilot$patid
time <- pilot$time / 12 # months -> years
status <- pilot$status
# # Baseline parameters for the Gumbel-Hougaard copula
gum <- gumbel.est(id, time, status)
lambda_D <- gum$lambda_D
lambda_H <- gum$lambda_H
kappa <- gum$kappa
tau <- 4 # 4 years of follow-up max
tau_b <- 3 # 3 years accrual
lambda_L <- 0.05 # loss to follow-up rate
bparam <- base(lambda_D, lambda_H, kappa, tau_b, tau, lambda_L)
# # Sample size (HR = 0.9 and 0.8 for death and hospitalization, respectively)
# # # 80% power, two-sided alpha = 0.05
WRSS(
  xi = log(c(0.9, 0.8)),
  bparam = bparam,
  q = 0.5,
  alpha = 0.05,
  power = 0.8
)$n
# # # 90% power, two-sided alpha = 0.05
WRSS(
  xi = log(c(0.9, 0.8)),
  bparam = bparam,
  q = 0.5,
  alpha = 0.05,
  power = 0.9
)$n

# Joint frailty model sample size ---------------------------------------------
library(frailtypack)

# # Assumptions
# # # - Accrual duration: 3 years
# # # - End of study: 4 years (3 years of accrual + 1 additional minimum year of follow-up)
# # # - Allocation ratio experimental:control of 1:1
# # # - Mean number of 3 recurrent events per patient in the control group (~Poisson distribution)
# # # - Median time to first hospitalization in the control group: 3.6 months
# # # - Median time to death in the control group: 28 months
# # # - The hazard increases over time (shapes > 1),
# # #   faster for death (shape = 2) than for hospitalization (shape = 1.5)
# # # - Hazard ratio for first hospitalization: 0.8
# # # - Hazard ratio for death: 0.9
# # # - Between-event correlation assumed (frailty variance = 1)
# # # - Recurrent events are positively associated with the terminal event (alpha = 1)
# # # - Bilateral joint test, with a two-sided alpha = 0.05

# # # # 80% power
JFM.ssize(
  power = 0.8,
  ni = 3,
  ni.type = "Pois",
  Acc.Dur = 3,
  FUP = 4,
  FUP.type = "UpToEnd",
  medianR.H0 = 3.6 / 12,
  medianD.H0 = 28 / 12,
  betaR.H0 = log(1),
  betaR.HA = log(0.8),
  betaD.H0 = log(1),
  betaD.HA = log(0.9),
  shapeR.W = 1,
  shapeD.W = 1,
  theta = 1,
  alpha = 1,
  ratio = 1,
  samples.mc = 1e5,
  seed = 42,
  timescale = "calendar",
  betaTest.type = "joint",
  statistic = "Wald",
  typeIerror = 0.05,
  test.type = "2-sided"
)

# # # # 90% power
JFM.ssize(
  power = 0.9,
  ni = 3,
  ni.type = "Pois",
  Acc.Dur = 3,
  FUP = 4,
  FUP.type = "UpToEnd",
  medianR.H0 = 3.6 / 12,
  medianD.H0 = 28 / 12,
  betaR.H0 = log(1),
  betaR.HA = log(0.8),
  betaD.H0 = log(1),
  betaD.HA = log(0.9),
  shapeR.W = 1,
  shapeD.W = 1,
  theta = 1,
  alpha = 1,
  ratio = 1,
  samples.mc = 1e5,
  seed = 42,
  timescale = "calendar",
  betaTest.type = "joint",
  statistic = "Wald",
  typeIerror = 0.05,
  test.type = "2-sided"
)

# Win ratio sample size through simulations -----------------------------------

# # Same assumptions as above

# Please install the modified WR package from the GitHub repository provided in Section 6. 
# The new function "sz_lwr()" has been added in this new version. You have two main ways to 
# do this:
# - You can either download "WR_1.0.tar.gz":
#   https://github.com/adrienorue/comparison_wrRec_jfm/raw/main/WR_1.0.tar.gz
# and install it locally using the following command in R:
# install.packages("path/WR_1.0.tar.gz", repos = NULL, type = "source")

# - Or you can follow these steps:
# # 1. Remove any existing CRAN version: remove.packages("WR")
# # 2. Install the "remotes" package (if not already available): install.packages("remotes")  
# # 3. Install the modified version: 
# #    remotes::install_url("https://github.com/adrienorue/comparison_wrRec_jfm/raw/main/WR_1.0.tar.gz")
# # If installation fails, please verify that "Rtools" is installed on your system.

library(WR) # modified version of WR package

res <- sz_lwr(
  power = 0.80,
  type1Error = 0.05,
  weibShapeRec = 1,
  weibScaleRec = (3.6 / 12) / log(2),
  weibShapeTerm = 1,
  weibScaleTerm = (28 / 12) / log(2),
  HR_recurrent = 0.70,
  HR_terminal = 0.80,
  niType = "poisson",
  ni = 3,
  theta = 0.5,
  alpha = 1.0,
  fupType = "uptoend",
  FUP = 4,
  accrualDuration = 3,
  sample_size_min = 1000,
  sample_size_max = 1400,
  sample_size_step = 100,
  n_sim = 2000,
  baseSeed = 42,
  output_plot_file = sprintf(
    "power_curve_%s.png",
    gsub("[ :\\-]", "_", round(Sys.time(), 0))
  )
)

\end{lstlisting}

\subsection{Master script}\label{app:master_script}
\begin{lstlisting}[
  language=R, 
  caption={Master script to reproduce all analyses}, 
  commentstyle=\ttfamily\upshape, 
  keywordstyle=\ttfamily\mdseries,
  basicstyle=\fontsize{8}{10}\selectfont\ttfamily
]
library(here)

# ------------------------------------------------------------
# ========== Simulations

# Generate scenario datasets; this may take some time
source(here("Simulations", "Generation.R"), print.eval = TRUE)


# For the "wrTrue" vector (lines 28 to 31), the true WR values for the "BIG" datasets are required.
# - Values are presented in the paper, Table 4, and in the code below (lines 34-35)
# - To reproduce them, uncomment the code below and run it (adapt the dataset path accordingly)
#   Caution: this may take some time.

# library(WR) # please install our optimized version of WR for this exact purpose
# source(here("Simulations", "helpers", "estimationWR.R"))
# trueWR_mc_run <- runMC_WR(
#     seedMax = 50,
#     seedStart = 1,
#     dataPath = here("Simulations", "datasets scenario 1BIG"), # adapt the path
#     id = "id",
#     time = "time",
#     status = "status",
#     trt = "z1",
#     # strata = "z2", # uncomment for stratified ; comment for unstratified
#     trueWR = 1
# )

# Change the dataset path according to the folders created by Generation.R,
# namely "datasets scenario X", for X = 1,2,3,4, 5, or 6
# Values for each scenario are presented in the paper, Table 4. For ease of use,
# here are the values for scenarios 1, 2, 3, 4, 5, and 6, respectively:
# - unstratified : 1.2253, 1.2901, 1.1912, 1.1962, 1.3465, 1.2514
# - stratified   : 1.2254, 1.2904, 1.1912, 1.1962, 1.3470, 1.2515
wrTrue <- c(
    unstratified = 1.2253,
    stratified = 1.2254
)
simulationDataPath <- "datasets scenario 1"
source(here("Simulations", "Description.R"), print.eval = TRUE)
source(here("Simulations", "Estimation.R"), print.eval = TRUE)

# ------------------------------------------------------------
# ========== Applications

source(here("Analyses", "HFaction analysis.R"), print.eval = TRUE)
source(here("Analyses", "Readmission analysis.R"), print.eval = TRUE)

#if TRUE, perform sample-size calculation using the simulation-based approach for the win ratio.
run_wr_simulation <- FALSE # if TRUE, this may take some time
source(here("Analyses", "Sample size based on HF-action.R"), print.eval = TRUE)

\end{lstlisting}

\subsection{Sample size derivation using Schoenfeld's approximation}\label{app:schoenfeld}

In the HF-ACTION study design paper \cite{Whellan2007-hk}, an annual event rate of 30\% was assumed in the control group. Under an exponential model for the time to the first event, this corresponds to a constant hazard rate
$$
  \lambda_C = -\log(1-0.30) \approx 0.3567\ \text{year}^{-1},
$$
which yields a 2-year event probability in the control arm of
$$
  \mathrm{P}_C(2) = 1 - \exp(-2\lambda_C) = 0.51.
$$

It was also assumed a net 11\% reduction in the 2-year event rate in the treatment arm, after accounting for non-adherence and crossover. Denoting by $\mathrm{P}_T(t)$ and $\mathrm{P}_C(t)$ the event probabilities at time $t$ in the treatment and control arms, respectively, this assumption may be written as
$$
  \mathrm{P}_T(2) = (1-0.11)\mathrm{P}_C(2) = 0.89 \times 0.51 = 0.4539.
$$
Under exponential hazards, this implies a constant treatment hazard rate
$$
  \lambda_T = -\frac{1}{2}\log\left(1-\mathrm{P}_T(2)\right)
  = -\frac{1}{2}\log\left(1-0.89[1-\exp(-2\lambda_C)]\right)
  \approx 0.3025\ \text{year}^{-1},
$$
hence leading to
$$
  \mathrm{HR} = \frac{\lambda_T}{\lambda_C} \approx 0.8480 \implies \log(\mathrm{HR}) \approx -0.1648.
$$

Patients were assumed to be enrolled uniformly over a 3-year accrual period, with a fixed study end at 4 years. Thus, if $U \sim \mathrm{Uniform}(0,3)$ denotes the calendar time of randomization, the individual follow-up duration is $T = 4 - U$, which ranges from 1 to 4 years. For arm $a \in \left(C,T\right)$ with exponential hazard $\lambda_a$, the probability that a patient randomized at time $U=u$ experiences at least one event before the administrative study end is
$$
  \Pr(\text{event} \mid U=u, a) = 1 - \exp\left(-\lambda_a(4-u)\right).
$$
Averaging over the uniform accrual distribution gives the average event probability in arm $a$:
$$
  \pi(\lambda_a) = \mathbb{E}\big[1 - \exp\left(-\lambda_a(4-U)\right)\big]
  = 1 - \frac{1}{3}\int_0^3 \exp\left(-\lambda_a(4-u)\right)du
  = 1 - \frac{\exp(-\lambda_a) - \exp(-4\lambda_a)}{3\lambda_a}.
$$
Evaluating this expression at the two arm-specific hazards yields
$$
  \pi_C = \pi(\lambda_C),
$$
and
$$
  \pi_T = \pi(\lambda_T).
$$
With 1:1 allocation, the average event probability across all randomized patients is therefore
$$
  \bar{\pi} = \frac{\pi_C + \pi_T}{2}.
$$

Using Schoenfeld's formula, we computed earlier a total of 1156 events for 80\% power and 1548 events for 90\% power. Finally, given the average per-patient event probability $\bar{\pi}$ under the planned accrual and follow-up, the corresponding total sample size is approximated by
$$
  N \approx \frac{d}{\bar{\pi}}.
$$
With $\bar{\pi} \approx 0.54$, we have:
$$
  N_{80\%} \approx \frac{1156}{\bar{\pi}} \approx 2132,
$$
and
$$
  N_{90\%} \approx \frac{1548}{\bar{\pi}} \approx 2856,
$$
after rounding up to the nearest even number to preserve equal randomization.

\end{document}